\documentclass[sigconf]{acmart}
\usepackage{soul} 
\usepackage{hyperref}
\usepackage{afterpage}
\usepackage{float} 
\newcommand{\zz}[1]{\textcolor{black}{#1}}

\AtBeginDocument{%
  }

\setcopyright{acmlicensed}
\copyrightyear{2025}
\acmYear{2025}
\acmDOI{XXXXXXX.XXXXXXX}

\acmConference[IUI'26]{Make sure to enter the correct
  conference title from your rights confirmation emai}{March 23--26,
  2026}{Paphos, Cyprus}
\acmISBN{978-1-4503-XXXX-X/18/06}

\copyrightyear{2026}
\acmYear{2026}
\setcopyright{cc}
\setcctype{by}
\acmConference[IUI '26]{31st International Conference on Intelligent User Interfaces}{March 23--26, 2026}{Paphos, Cyprus}
\acmBooktitle{31st International Conference on Intelligent User Interfaces (IUI '26), March 23--26, 2026, Paphos, Cyprus}
\acmDOI{10.1145/3742413.3789112}
\acmISBN{979-8-4007-1984-4/2026/03}

\begin{document}

\title{Gazeify Then Voiceify: Physical Object Referencing Through Gaze and Voice Interaction with Displayless Smart Glasses}

\author{Zheng Zhang}
\authornote{Work done during the author’s internship at Meta Reality Labs}
\orcid{0000-0002-7040-2326}
\affiliation{%
  \institution{University of Notre Dame}
  \city{Notre Dame}
  \state{IN}
  \country{USA}
}

\author{Mengjie Yu}
\affiliation{%
  \institution{Meta Reality Labs}
  \city{Redmond}
  \state{WA}
  \country{USA}
}

\author{Tianyi Wang}
\affiliation{%
  \institution{Meta Reality Labs}
  \city{Redmond}
  \state{WA}
  \country{USA}
}

\author{Kashyap Todi}
\affiliation{%
  \institution{Meta Reality Labs}
  \city{Redmond}
  \state{WA}
  \country{USA}
}

\author{Ajoy Savio Fernandes}
\affiliation{%
  \institution{Meta Reality Labs}
  \city{Redmond}
  \state{WA}
  \country{USA}
}

\author{Yue Liu}
\affiliation{%
  \institution{Meta Reality Labs}
  \city{Redmond}
  \state{WA}
  \country{USA}
}

\author{Haijun Xia}
\affiliation{%
  \institution{UC San Diego}
  \city{La Jolla}
  \state{CA}
  \country{USA}
}

\author{Tovi Grossman}
\affiliation{%
  \institution{University of Toronto}
  \state{Toronto}
  \country{Canada}
}

\author{Tanya Jonker}
\affiliation{%
  \institution{Meta Reality Labs}
  \city{Redmond}
  \state{WA}
  \country{USA}
}



\renewcommand{\shortauthors}{Zheng et al.}

\begin{abstract}

    Smart glasses enhance interactions with the environment by using head-mounted cameras to observe the user’s viewpoint , but lack the visual feedback used for common interactions. We introduce ``Gazeify then Voiceify'', a multimodal approach allowing object selection via gaze and voice using displayless smart glasses. Users can select a physical object with their gaze, and the system generates a digital mask and a voice description of the object's semantics. Users can further correct errors through free-form conversation. To demonstrate our approach, we develop an interactive system by integrating advanced object segmentation and detection with a visual-language model. User studies reveal that participants achieve correct gaze selection in 53\% of the task trials and use voice disambiguation to correct 58\% remaining errors. Participants also rated the system as likable, useful and easy to use.

  
\end{abstract}



\begin{teaserfigure}
\centering
\includegraphics[width=0.78\textwidth]{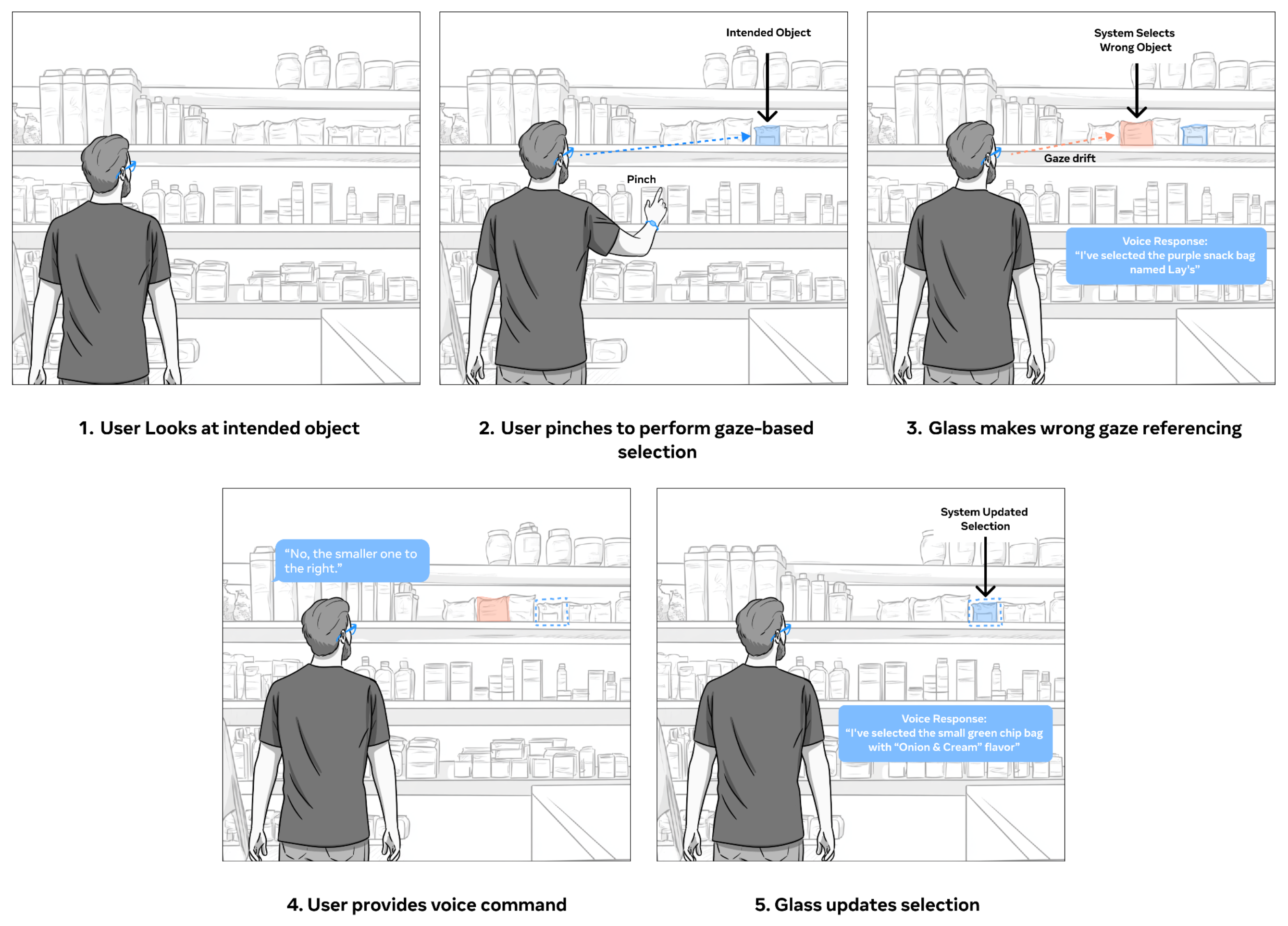}
\caption{Interaction flow with Gazeify Then Voiceify: Users initiate gaze-based object selection by looking at the target and pinching. The system then provides a description of the selected object. If a selection error is detected, users can issue a voice command to make corrections. The glasses will then describe the updated selection for confirmation.}
\label{fig:teaser_figure}
\end{teaserfigure}

\maketitle

\section{INTRODUCTION}

Displayless smart glasses, such as the Ray-Ban Meta smart glasses\footnote{\href{https://www.meta.com/smart-glasses/}{https://www.meta.com/smart-glasses/}} and Aria glasses\footnote{\href{https://about.meta.com/realitylabs/projectaria/}{https://about.meta.com/realitylabs/projectaria/}}, offer a lightweight and comfortable alternative to traditional head-mounted displays (HMDs), making them ideal for all-day wear. With an integrated camera-enabled AI assistant, current smart glasses have enhanced the convenience of AI-driven environmental interaction. For example, if a user wishes to know a competitor's price for a can of coffee in a grocery store, they can aim their glasses' camera at the coffee can and ask, \textit{"What is the price of this coffee in other stores?"} The AI assistant would then identify the target item using the data captured by the glasses' camera. However, this interaction largely depends on the AI's ability to accurately navigate scene ambiguity and recognize the intended object in the environment. 

Previous research on gaze-based object selection within XR environments has shown the efficiency of gaze selection over alternative methods like head pointing~\cite{Sidenmark2024ConeBubbleEC, Angelo1991ComparingTH, fernandes2023leveling}, voice~\cite{Miniotas2006SpeechaugmentedEG} and gestures~\cite{Jang2017MetaphoricHG, Bhowmick2021UnderstandingGP, Ren20133DSW, squiggle2025}. Recent systems such as WorldGaze~\cite{Mayer2020EnhancingMV} and GazeGPT~\cite{konrad2024gazegpt}, have thus incorporated gaze tracking to better align a system's focus with a user's attention. While gaze tracking can improve reference accuracy in simple settings, it can introduce errors, such as saccades during queries, which prevents it from performing well in cluttered environments. This exemplifies the need for disambiguation methods to correct potential gaze errors. Prior work has explored using input such as head~\cite{sidenmark2019eye, sidenmark2020bimodalgaze}, voice~\cite{Bolt1980PutthatthereVA} and hand gestures~\cite{kyto2018pinpointing, istance2010designing} to perform post-hoc disambiguation. These methods, however, typically assume that visual feedback is present so users can continually understand the selection progress. Moreover, these techniques have mainly been applied to the selection of virtual objects, where objects possess well-defined geometric and spatial characteristics. To overcome the obstacles associated with the absence of visual feedback, previous studies~\cite{Cho2024SonoHapticsAA} have employed sensory feedback mechanisms like acoustic and haptic feedback to represent the visual characteristics of objects. These approaches typically require the use of additional devices, such as smartwatches. Moreover, they depend on pre-defined cross-modal mappings. When trying to select objects in real-world environments, the variety of visual cues is often dynamic and may not consistently correspond to these predefined sensory feedback mechanisms.

Compared to other feedback methods, voice has a potential to provide a more natural and accessible means for selection disambiguation when using smart glasses without displays. For an ideal voice interaction workflow, smart glasses should deliver responses that include specific semantic features of the selected object, such as its relative position, color, and category to help users understand the results of their selection. Conversely, users should have the flexibility to issue voice commands to correct a selection if they determine that is incorrect. Recent advancements in Vision Language Models (VLMs)~\cite{Sonoda2024DiagnosticPO, Zhang2023VisionLanguageMF} have demonstrated how such models can comprehend visual scene and element layouts, enabling on-the-fly vocal descriptions of selection results. The language-driven correction of object segmentation remains a challenge, however, as these models still have limited capabilities to understand and respond to natural language input~\cite{Liu2023GroundingDM, Xu2023BridgingVA}.

Amidst these opportunities and challenges, we present a new interaction technique for smart glasses called \textbf{physical object referencing}. This technique enables users to quickly select real-world objects using their gaze, allowing actions on their digital masks derived from gaze-based segmentation. Additionally, since gaze selection is prone to noise, a method is needed to help users recognize and correct errors. Leveraging the benefits of gaze and voice input, we created "Gazeify Then Voiceify" (Figure \ref{fig:teaser_figure}), a technical probe that enables gaze-based physical object referencing and voice-based follow-up disambiguation for displayless smart glasses. With Gazeify Then Voiceify, users can simply look at a target object and trigger a selection with a pinch or button click. Then, Gazeify will conduct gaze-driven object segmentation and generate a digital mask of the object. Voiceify will then inform the user about the selection result via audio descriptions generated by a vision language model (VLM). If there are selection errors, the user can utilize free-form voice commands to fix them.

A user study was conducted to assess the effectiveness, usability, and user experience when using "Gazeify Then Voiceify" when factors like the perceived size of objects, environmental clutter, and structural and positional object ambiguities were manipulated. The findings showed that participants encountered more challenges when attempting to select small objects, large objects with complex structures, and objects in cluttered environments. Participants were also able to use the audio feedback to determine the accuracy of a selection. However, Voiceify tended to produce incorrect or ambiguous audio descriptions when faced with multiple similar objects in close proximity or when the object was small in the field of view. Lastly, Voiceify can correct the majority of gaze selection errors through user commands when the object is sufficiently clear for detection. The technique was also found to be user-friendly, low-friction, and provided a positive user experience. 

Our research thus contributes:

\begin{itemize}
    \item A gaze-driven physical object referencing technique that enables users to select and generate a mask of object of interest simply using their gaze.
    \item A VLM-based approach that enables users to identify and disambiguate selection errors using free-form voice feedback.
    \item A controlled user study that identified, among other findings, the role that the perceived size of objects, environmental clutter, and structural and positional object ambiguities have on gaze referencing, voice description and disambiguation accuracy.
\end{itemize}

\section{RELATED WORK}
In this section, we review the existing literature related to eye gaze interaction, physical object interaction with head-mounted devices, and techniques for object detection and segmentation.

\subsection{Eye Gaze Interaction}

Eye gaze-based interaction offers users speed, convenience, and direct availability in AR/VR environments~\cite{tanriverdi2000interacting, Shi2024CasualGazeTM}. Compared to other input methods such as head pointing or ray casting, eye gaze has been found to be less strenuous~\cite{blattgerste2018advantages, hansen2018fitts, kyto2018pinpointing, qian2017eyes, fernandes2023leveling}. This enables it to offer users a more fluid, intuitive interaction experience. 

Eye gaze-based interaction can be classified as unintentional (implicit) or intentional (explicit)~\cite{majaranta2014eye, sendhilnathan2024implicit}. Unintentional gaze interaction tracks the user's gaze to infer their focus, attention, or intent without the user being fully aware of it. Such pervasive tracking of gaze has enabled for a range of context-aware applications such as adaptive interfaces~\cite{majaranta2014eye, nielsen1993noncommand, sendhilnathan2024implicit}, information management~\cite{toyama2014natural, giannopoulos2015gazenav}, and question answering~\cite{kwok2019gaze, lee2024gazepointar, konrad2024gazegpt}.

On the other hand, with intentional gaze interaction, users intentionally focus their gaze on a specific object or area to perform an action. At doing so, they can perform gaze-only interactions such as typing~\cite{rajanna2018gaze}, object augmentation~\cite{bace2016ubigaze}, or object selection~\cite{piumsomboon2017exploring}, or multimodal operations like manipulating gaze-selected object with free-hand gestures~\cite{deng2017understanding, pfeuffer2017gaze+, schweigert2019eyepointing}. Despite the naturalness of eye gaze, intentional gaze interaction suffers from low accuracy due to calibration offsets and noise, and drift within eye-tracking sensors~\cite{cesqui2013novel, larsson2016head, piumsomboon2017exploring, fernandes2024degraded}. Furthermore, it is also vulnerable to the "Midas Touch" problem~\cite{jacob1990you}, where unintended actions are triggered by a user's natural eye movements, resulting in incorrect selections and reduced usability. Researchers have investigated a variety of solutions to improve intentional gaze interaction accuracy with virtual objects, such as using dwell time~\cite{jacob1990you, park2008wearable, sibert2000evaluation}, smooth pursuits~\cite{sidenmark2020outline, esteves2017smoothmoves, khamis2017eyescout, vidal2013pursuits}, gaze gestures~\cite{bace2016ubigaze, drewes2007interacting, hyrskykari2012gaze, istance2010designing} and movement patterns~\cite{sidenmark2023vergence, lu2021exploration, pai2016transparent}. In addition, additional modalities such as head pointing~\cite{sidenmark2019eye, sidenmark2020bimodalgaze, wei2023predicting} and hand gestures~\cite{istance2010designing, kyto2018pinpointing} have been proposed to disambiguate gaze interaction in extended reality.

While prior research has focused on the precise selection of virtual objects using eye gaze, the selection of real-world objects using eye gaze is underexplored. The present work presents a segmentation based techniques that enables user to perform gaze-driven selection of real-world objects, and obtain a digital mask of the object for further use.


\subsection{Physical Object Interaction with Head-Mounted Devices}

Prior research has explored various ways to interact with real-world objects through smart head-mounted devices. Targeting digital content augmented on physical object, Dogan et al. classified physical object interaction in Extended Reality (XR) across two dimensions: anchoring (i.e., manual vs. seamless) and content (i.e., arbitrary vs. object focused; ~\cite{Dogan2024AugmentedOI} ). The manual anchoring approach required the pre-registration of objects and manual setups to achieve tangible input~\cite{CamposZamora2024MoirWidgetsHP, Ruofei2022OpportunisticIF, Monteiro2023TeachableRP}. In contrast, seamless anchoring automatically detected the object and executed AR content anchoring~\cite{Dogan2024UbiquitousMD, Chen2020AugmentingSV, Henderson2008OpportunisticCL}. In terms of digital content, some research has focused on using tangible interaction with physical objects~\cite{Monteiro2023TeachableRP, Suzuki2020RealitySketchER} to control digital content or create digital twins~\cite{Chidambaram2022EditARAD}. Other work like XR-Objects~\cite{Dogan2024AugmentedOI}, InfoLED~\cite{Yang2019InfoLEDAL}, and Reality Editor~\cite{Heun2013RealityEP} enabled object-specific content, such as inferring actions that objects can afford, to enhance dynamic physical object interaction.

Apart from the digital augmentation of content on physical objects, recent advances in AI have enabled smart head-mounted devices to seamlessly answer user's queries about the physical environment. For example, enabled users to ask questions about physical objects using pronoun references and leveraged AI to automatically disambiguate the referring object in the context GazePointAR~\cite{lee2024gazepointar}. To better align a device's focus and user's attention, work like GazeGPT~\cite{konrad2024gazegpt}, Gaze-guided Narrative~\cite{kwok2019gaze} and WorldGaze~\cite{Mayer2020EnhancingMV} incorporated user gaze information during queries and adapted auditory feedback based on gaze changes. Other work has also combine voice queries with touch~\cite{Lee2021WhatsTA} or pointing~\cite{Romaniak2020NimbleMI, konrad2024gazegpt} to clarify the object a user wants to interact with. \zz{Furthermore, recent research has expanded beyond explicit object querying to explore broader everyday assistance specifically within lightweight smart glasses. For instance, Cai et al. introduced AiGet to facilitate hidden knowledge discovery during everyday moments \cite{cai2025aiget}, while Pu et al. proposed ProMemAssist to offer timely proactive assistance by modeling working memory in multi-modal wearable devices \cite{pu2025promemassist}. These works highlight the growing utility of smart glasses for continuous, ambient support, distinguishing them from the interaction paradigms of traditional HMDs.}

Unlike prior work on augmenting or querying physical objects, we focus on selecting physical objects with smart glasses that are limited in their ability to provide visual feedback. Compared with prior work~\cite{konrad2024gazegpt, lee2024gazepointar} utilizing VLMs for question answering, we leveraged VLMs to locate and describe objects in context and support language-driven mask updating. 


\subsection{Object Detection and Segmentation}

\subsubsection{Object detection technology}

Object detection involves predicting the category tags and bounding box coordinates for each object in an image. Current methodologies can be categorized into two types: two-stage and one-stage detectors. Two-stage detectors, such as RCNN~\cite{girshick2014rich} and its derivatives~\cite{cai2018cascade, he2017mask, ren2016faster}, initially extracted a set of region proposals and performed classification and regression on these regions. On the other hand, one-stage detectors~\cite{lin2017focal, liu2016ssd, redmon2016you} bypassed the proposal stage and directly classify results for predefined anchors. Additionally, transformer-based methods ~\cite{carion2020end, dai2021up, zhang2022dino} have been rapidly advancing in this field. 

Despite these advancements, most traditional methods operate using a closed-world assumption~\cite{joseph2021towards}, where they can only detect categories present in their training datasets. To overcome this limitation, open-vocabulary object detection has emerged~\cite{zhao2023revisiting, bansal2018zero, zhu2019zero}. Recent developments have leveraged large-scale, image-text pre-training to enhance open-vocabulary detection~\cite{du2022learning, feng2022promptdet, gu2021open}, significantly expanding performance and category recognition capabilities. Innovations such as UniDetector~\cite{wang2023detecting} have furthered generalizability by training models across multiple image sources and diverse label spaces to align image and text representations. Current state-of-the-art models ~\cite{zhu2024survey, wang2023detecting, awais2023foundational} may, however, fail to recognize everyday objects in complex, natural environments.

\subsubsection{Object segmentation technology}

Object segmentation aims to produce pixel-level masks of objects, delineating their precise boundaries within an image. Image segmentation encompasses several sub-disciplines, including instance segmentation, semantic segmentation, and panoptic segmentation~\cite{he2017mask, chen2017deeplab, kirillov2019panoptic}, with each targeting varying levels of semantic detail. For instance, semantic segmentation classifies each pixel according to its semantic class~\cite{chen2017rethinking, long2015fully}, while instance segmentation not only classifies each object instance but also differentiates individual object instances within the same class~\cite{he2017mask, li2023mask}. Recent works have explored interactive segmentation~\cite{xu2016deep, kirillov2023segment, Liu2022SimpleClickII, Zou2023SegmentEE}. For example, Segment Anything (SAM)~\cite{kirillov2023segment} can utilize user input such as points, boxes, or scribbles to segment objects of interest. 

\paragraph{Gaze-based segmentation} Some Machine Learning works have explored to use gaze as prompt for  initiating object segmentation. For example, SAM meets Gaze~\cite{Beckmann2023SAMMG} and GazeSAM~\cite{Wang2023GazeSAMII} have leveraged passive eye-tracking to gather gaze data for annotating point-prompted segmentation datasets efficiently. These efforts, however, primarily utilize pervasive gaze data for accelerating segmentation model training rather than enabling users to actively use gaze for object segmentation in real-world scenarios.  

\paragraph{Language-driven segmentation} More recent studies have investigated language-driven segmentation~\cite{Liu2023GroundingDM, Xu2023BridgingVA, Wu2020PhraseCutLI, rasheed2024glamm}. However, these models typically accept only short phrases (e.g., object types) as input and struggle with generalization to unseen categories. They also exhibit limited understanding of common referential language, particularly expressions involving spatial, ordinal, and part-whole relationships. Additionally, while VLMs have shown proficiency in image understanding and creation, they still struggle with precise object detection and segmentation tasks. This limitation stems from insufficient training on these tasks and the loss of precise coordinate information during image processing and transmission to VLMs. Therefore, this area remains largely unresolved. 

In this work, we present an interactive technical pipeline that allows users to segment reference objects using gaze, with voice commands for refining the results. We have developed a retrieval-based approach for proxying language-driven segmentation, which leverage a VLM to identify the most likely object mask based on user command from a set of candidates produced through global segmentation.

\section{GAZEIFY THEN VOICEIFY SYSTEM DESIGN}

\begin{figure*}[ht]
\centering
\includegraphics[width=\textwidth]{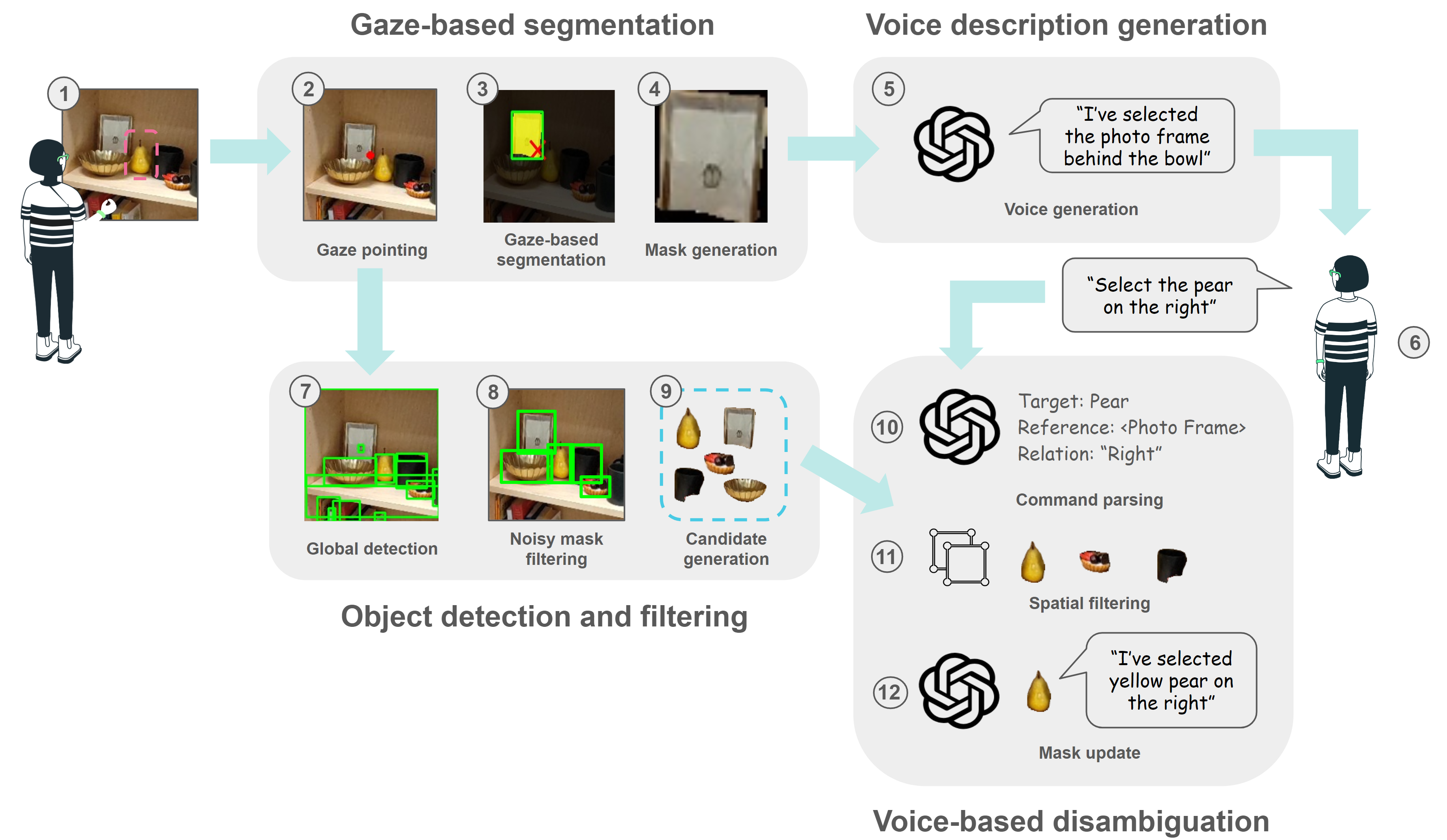}
\caption{The interaction flow and system architecture illustration. (1) User targets object and activates selection with a pinch; (2) User's gaze unintentionally shifts off target, the red dot represents gaze; (3) Gazeify incorrectly segments an object; (4) Gazeify outputs an incorrect object mask; (5) Voiceify provides a description of the selection; (6) User detects the error and issues a voice command for correction; (7) Upon gaze selection activation, the system identifies comprehensive objects in the focal view; (8) VLM filters out noisy detections; (9) Candidate object masks are generated and sent to Voiceify; (10) VLM processes the user's command; (11) Heuristic filtering uses spatial relations identified by VLM; (12) VLM selects the most likely object from candidates based on the user command and notifies the user.}
\label{fig:system_architecture}
\end{figure*}

This section introduces Gazeify Then Vocieify technique (Figure \ref{fig:system_architecture}). Next, we describe its gaze-based object referencing, voice description generation, voice command comprehension, object mask updating, and implementation.
 
\subsection{Gaze-Based Physical Object Referencing}
\label{sec:gaze_selection}
Gazeify allows users to generate a mask of an object that they are looking at using a simple pinch gesture or button click (Figure \ref{fig:system_architecture}.1). Once they have pinched or clicked, an EfficientSAM model~\cite{Xiong2023EfficientSAMLM} extracts visual information from the color camera image captured and transforms it into an image embedding.
Simultaneously, Gazeify employs a novel \textit{spatiotemporal-aware gaze sampler} to determine the gaze points to use as a point-based prompt to drive object segmentation. The sampler processes both the image and gaze points sampled within a time window, $\delta$, around the selection. It extracts their features and clusters the gaze points based on the similarity of the spatial and visual features. We used four key, gaze-based features: color, gaze depth, location, and gaze velocity. The color feature helps cluster gaze points on similar objects, based on the assumption that similar objects have similar colors. The gaze depth feature differentiates between objects at different distances. The location feature clusters gaze points that are close together in space, based on the assumption that gazes close to each other have similar goals. The gaze velocity feature clusters gaze points exhibiting similar behaviors, like fixations or saccades. After creating the clusters, the model identifies the largest cluster, which is assumed to represent the area with the most attention in recent history. The model then calculates the centroid of the cluster, which is then passed to EfficientSAM's prompt encoder to retrieve its embedding.

After the image and gaze embedding are calculated, they are passed to EfficientSAM's mask decoder. The model then outputs three masks of varying granularities, each with a corresponding confidence score. Gazeify selects the mask with the highest confidence score as the output mask for object referencing (Figure \ref{fig:system_architecture}.4).

\subsection{Voice Description Generation}
\label{sec:voice_description}

Voiceify generates voice descriptions (Figure \ref{fig:system_architecture}.5) that inform users about the semantic objects within the gaze selection mask (Section \ref{sec:gaze_selection}) or the updated mask resulting from voice disambiguation (Section \ref{sec:voice_disambiguation}). The description enables users to identify the currently selected objects in their physical surroundings and identify any selection errors. To achieve this, we leveraged VLM (GPT-4o\footnote{\href{https://openai.com/index/hello-gpt-4o/}{https://openai.com/index/hello-gpt-4o/}}) to understand the mask semantics and describe the masked object based on its spatial relationships to nearby objects and distinguishable characteristics.

Once the resulting mask is obtained, Voiceify provides the VLM with both the mask and a cropped context image centered at the centroid of the sampled gaze points that were used for initial referencing. The cropped image corresponds to the tight bounding box around the mask, with an additional padding of $\sigma$. This cropping helps the VLM focus on the specific region that the user was paying attention to, ensuring that the generated description emphasizes the immediate surroundings of the selected object. Then, Voiceify prompts the VLM to generate a voice description by locating the masked object within the context image. The description follows the template: "\textit{I've selected [adjective properties] [object identity] that [spatial/ordinal relationship to nearby objects]}." Specifically, the adjective properties refer to characteristics of the masked object, such as color, size, shape, texture, and material. The object identity describes what the masked object is, such as a cup, headphones, or a chair. If the mask shows part of an object, e.g. "cap of a bottle" or "logo of a snack bag", the VLM must clarify this belonging relationship. Finally, the VLM distinguishes the masked object from nearby anchoring objects based on spatial or ordinal relationships. Spatial relationships include terms like left, right, above, below, behind, in front of, and between (e.g., "the cup to the left of the gold album" or "the beverage can between the mouse and the book"). When multiple similar objects (e.g., a row of bananas) are present in the context, the VLM describes the ordinal relationship of the masked object amongst them, such as "the middle beverage can among the three" or "the leftmost pumpkin". 

\subsection{Voice Command Comprehension}
\label{sec:command_comprehension}

\begin{figure*}[ht]
\centering
\includegraphics[width=\textwidth]{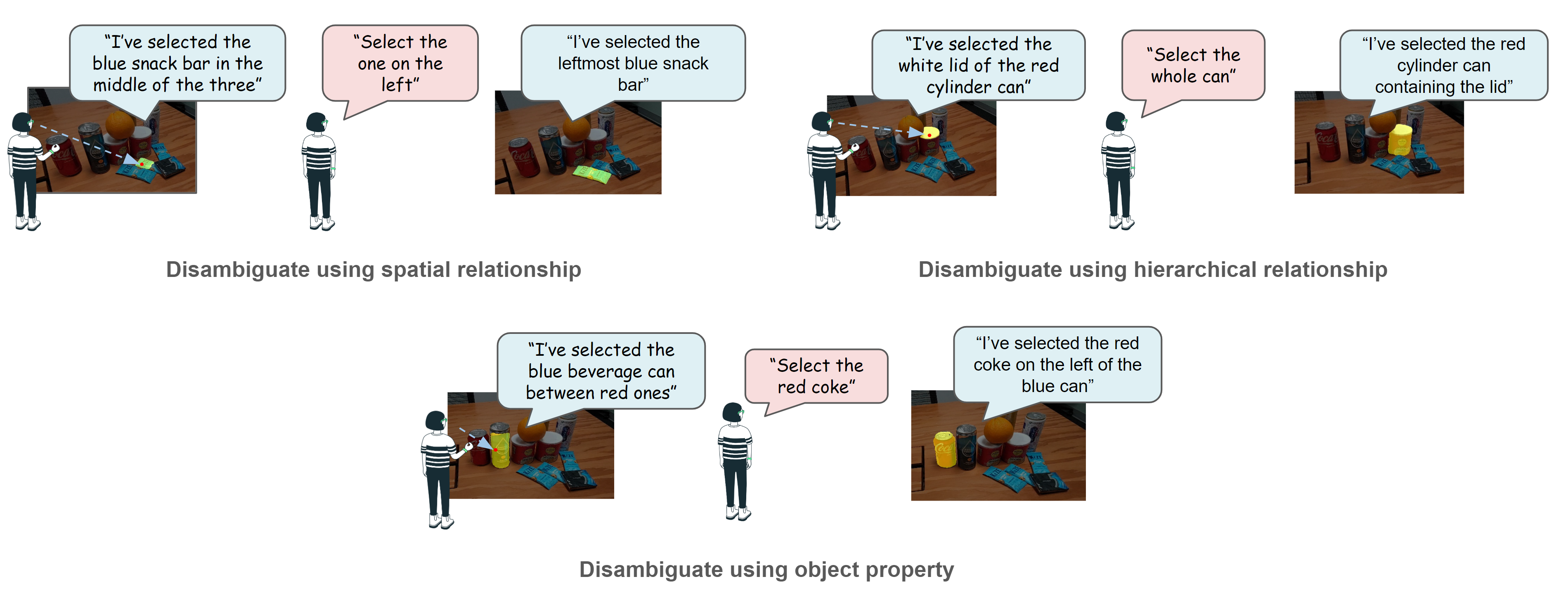}
\caption{The illustration of different disambiguation strategies that users can utilize. Users can specify object property, hierarchical, spatial (ordinal) relationships for instructing the system to correct selection.}
\label{fig:voice_command}
\end{figure*}

The Voice Command Comprehension module parses the user's free-form voice commands based on the context of the dialog history, identifying the target object, potential reference object, and the relationship between them (Figure \ref{fig:system_architecture}.10). This information is then utilized to filter candidate object masks and identify the target objects amongst the filtered objects (Section \ref{sec:obj_filtering}).

Voiceify enables users to employ common relative referential strategies, including specifying the target object's spatial relationships (e.g. left, right, above, below, behind, in front of), ordinal relationships (e.g. rightmost, leftmost, middle, [ordinal number]), or belonging relationships (e.g. part of, include) to reference objects (Figure \ref{fig:voice_command}). Upon receiving a user command, the VLM identifies the target object, reference object, and their relationships. The VLM then tries to map these relationships to one of the aforementioned types. If no explicit relationship is mentioned in the user's command (e.g. "the red cup"), the VLM will default to setting the relationship as "next to", assuming that the target object is in close proximity to the object described by the system during the previous selection. For both the target and reference objects, the VLM includes any adjective properties mentioned by the user, which can aid in later object localization to find an intended object.

The conversation history between the user and the system, along with the context image, helps Voiceify to resolve pronouns or implicit references in the user's command. By sending the conversation history to the VLM while parsing the command, we instructed the VLM to identify pronouns and implicit reference objects in the command and recover their references based on the past conversation. For example, if the system described the object selected during the last selection as "a blue beverage can between two bottles" and the user gives the command "select the red one to the left (of the blue can)", the VLM is expected to resolve "red one" to "the red beverage" in the context image and infer that the blue can is the reference object. We only instruct the VLM to infer the implicit reference object when an explicit relationship is provided but the reference object is missing in the user command.

\subsection{Object Mask Update Pipeline}
\label{sec:voice_disambiguation}

Voiceify employs a three-stage pipeline to update the mask to the target object in the parsed voice command. The pipeline consists of object detection, bounding box filtering, and object mask localization. The pipeline addresses the limitations of existing language-driven, open-set object detection and segmentation models.

\subsubsection{Object Detection}

The goal of the Object Detection stage is to identify as many objects as possible within the context image that presents focal area of user interest (Section \ref{sec:voice_description}). It enables Voiceify  to search for the user-described target object among the detected candidates during the Object Localization stage (Section \ref{sec:object_localization}). Initially, we experimented with multiple open-set detect everything models, including Grounding DINO~\cite{Liu2023GroundingDM}, OWLv2~\cite{Minderer2023ScalingOO}, and RTDETR~\cite{Lv2023DETRsBY}. However, we found that these models tended to miss a significant portion of objects in egocentric context images captured in open-world scenarios such as offices and kitchens. 

To enhance the detection of objects, we developed a method that integrates a global segmentation mode of Segment Anything Model (SAM) with the open-set detect anything model, RTDETR. For each mask resulting from the gaze referencing, we compute a tight bounding box within the cropped context image. Concurrently, the RTDETR model detects and create bounding boxes for other objects within the same context image. Upon obtaining bounding boxes from both sources, we employ a non-maximum suppression algorithm~\cite{Neubeck2006EfficientNS} on the combined set of bounding boxes to eliminate duplicate object detections (Figure \ref{fig:system_architecture}.7).

\subsubsection{Bounding Box Filtering}
\label{sec:obj_filtering}
After receiving candidate object bounding boxes, the bounding box filtering stage eliminates any noisy or irrelevant detections to reduce the search space for the subsequent Object Localization stage. First, Voiceify leverages GPT-4o to identify and remove noisy bounding boxes (Figure \ref{fig:system_architecture}.8). We categorized a noisy bounding box as one that emphasized insignificant parts of an object, the background, or distinctly separate objects. Additionally, we supplied GPT-4o with a context image to facilitate the analysis of image patch shown in each bounding box.

Then, Voiceify filtered out irrelevant bounding boxes that did not have the correct relationship to the reference object as detected in the Voice Command Comprehension (Section \ref{sec:command_comprehension}). As it was unclear if GPT-4o could process spatial filtering, we experimented with two methods to specify the model with the spatial relationships among bounding boxes, i.e., (i) overlaying indexed bounding boxes on the context image, and (ii) supplying the context image with each bounding box as an individual image patch with their top-left and bottom-right coordinates. Unfortunately, with the first method, the model struggled to accurately identify the index of each box from the image. With the second method, GPT-4o often confused the spatial relationships among the boxes, leading to incorrect filtering.

Therefore, we developed a novel heuristic method. First, we identified the bounding box for the reference object. If a reference object was specified in the Voice Command Comprehension, we used the tight bounding box from the object mask in the previous selection as the default reference box. We then prompted GPT-4o to assess whether the box matched the semantic description of the reference object derived from the command. If there was a discrepancy, the model was instructed to determine the most probable bounding box for the reference objects from the provided image patches. After identifying the reference bounding box, we applied a set of heuristics to each bounding box to verify if it maintained the specified spatial relationship to the reference box (Figure \ref{fig:system_architecture}.11). A bounding box was considered to be on the correct side of the reference if it was completely on the specified side of the reference, or if it overlapped with the reference box but the ratio of the overlapping margin to the respective side length of the reference box was below a certain threshold, $\alpha$. We experimentally determined $\alpha$ to be 0.5. For the relationship type "next to," we retained the seven bounding boxes that had the shortest distance between their center and the centroid of sampled gaze points or center of the reference object if it was specified.

\subsubsection{Object Localization}
\label{sec:object_localization}
In the final stage, we utilized GPT-4o to identify the target object among the filtered bounding boxes. We provided the model with the parsed command, individual image patches cropped from each bounding box, the context image, and the centroid coordinates of each bounding box within the context image space. With this information, we instructed the model to evaluate the likelihood that each image patch represented the described target object, explain their reasoning for each patch, and assign a relative confidence score ranging from 0 to 1. For ordinal relationship types, we requested the model identify multiple patches that matched the target object description, deduce their ordinal relationships using the centroid coordinates of their bounding box, and select the patch that correctly aligns with the specified ordinal position. In scenarios where multiple objects matched the description, we directed the model to prioritize the object nearest the reference object or to the centroid of the sampled gaze points if the reference was not provided.

We designated the patches with the highest score as the target object if their score exceeded 0.5. If the corresponding bounding box was derived from the Segment Anything model, we applied the corresponding mask as the updated mask. Otherwise, the EfficientSAM model executed box-based segmentation using the bounding box. Afterwards, the mask was forwarded to the voice description generation component to create a voice response. If the highest score was less than 0.5, we considered it a system failure. Nonetheless, we still directed GPT-4o to identify the most likely object from the context image and generate a voice command  (Figure \ref{fig:system_architecture}.12) ``\textit{Do you want to select [object description]?}'' where the description adhered to the template outlined in Section \ref{sec:voice_description}. This helped us to investigate in user study whether VLM could locate target object even during bounding box generation or filtering failures.

\subsection{Implementation}

\zz{We utilized the Meta Quest Pro as a hardware proxy for displayless smart glasses. Although the Quest Pro is traditionally categorized as an HMD, it was selected for this study due to its well-established development environment and integrated sensor suite (i.e., eye tracking, egocentric color camera, and microphone). To strictly simulate the constraints of the intended device class, we operated the device solely in passthrough mode and deliberately avoided rendering any immersive visualizations or augmented interfaces.}

Gazeify Then Voiceify was divided into front-end and back-end applications. The front-end application, developed in Unity, recognized pinch gestures or controller button clicks to activate gaze selection and also provided voice feedback. Additionally, Wit.ai\footnote{\href{https://wit.ai/}{https://wit.ai/}} was used for the speech recognition of user commands and for voice synthesis for the generated descriptions. The back-end application was built in Python and ran on a Windows PC equipped with an Nvidia 4080 GPU. This setup enabled for the use of local models for object detection and segmentation, including EfficientSAM, REDETR, and the Segment Everything Model. We employed GPT-4o-mini\footnote{Prompts of each key function are provided in Supplementary  Material} to handle all VLM tasks.

As the Quest Pro has a gaze sampling rate of 90 Hz, with raw color images sized at 1080 x 1080 pixels,  we utilized gaze points within a 0.5-second window before and after the selection time for the spatiotemporal-aware gaze sampler (Section \ref{sec:gaze_selection}). The padding for the context image (Section \ref{sec:voice_description}) was set to 150 pixels to ensure it captured nearby objects in most cases while excluding irrelevant areas.
\section{USER STUDY}
To assess the effectiveness, usability, and user experiences when using Gazeify Then Voiceify, we conducted a within-subject lab study. In this study, participants were asked to select different objects in a coffee shop-like setting, which had different object sizes, amounts of environmental clutter, object structural complexity and existence of similar objects. We sought to investigate the following research questions:

\begin{itemize}
    \item \textbf{RQ1}: How accurate is using gaze when selecting physical objects when their perceived size, the amount of environmental clutter, and object structural and positional ambiguities are manipulated?
    \item \textbf{RQ2}: To what degree can participants accurately determine what has been selected and detect gaze referencing errors based on the voice description?
    \item \textbf{RQ3}: To what degree can participants use voice commands to disambiguate gaze referencing errors they notice?
    \item \textbf{RQ4}: How usable Gazeify Then Voiceify and what is the overall user experience when using the technique? 
\end{itemize}

\subsection{Participants}
Twenty-three participants were recruited to participate in the study from an internal email list at a technical company and external participant pools (14 males, 8 females, and 1 individual who preferred not to disclose their gender). Regarding their experience with smart head-mounted devices such as AR/VR and smart glasses, 1 participant had never used such technologies, 2 had tried the technology once or twice, 10 used the devices several times, 6 were quite familiar with these technologies, and 4 were experts. Regarding their experience with generative AI, 1 participant had no prior exposure, 4 were aware of generative AI but not well-acquainted with it, 5 had used it on a limited basis, 10 frequently utilized the technology, and 3 were experts. The average eye calibration error of participants was $1.16^\circ$ ± $0.22^\circ$.

Participation was voluntary for the 10 internal participants. The study protocol was reviewed and approved by the IRB at our institution. 

\subsection{Trial Conditions}

Based on the everyday scenarios, we considered three factors of interest during the study, i.e., perceived object size, environmental clutter, and ambiguity. 

\begin{itemize}
    \item \textbf{Perceived Object Size}: Two perceived object sizes were used, i.e., \textit{small} and \textit{normal} (baseline). To ensure consistency in the perceived size of an object across the study we asked participants to stand on a designated marker on the floor while selecting a target object. The distance between the target object and the participant's head was maintained between 60 to 100 cm. We defined the perceived size of an object based on its visual angle from the perspective of a person who was approximately 170 cm tall and standing on the designated marker. Objects with a size less than $5^\circ$ x $5^\circ$ in visual angle were classified as small, while those with a larger visual angle were considered to have a normal size.

    \item \textbf{Environmental Clutter}: The environment used during the study as either \textit{cluttered} or \textit{clean} (baseline). In the cluttered condition,  the target object was occluded or surrounded in close proximity by other objects. Conversely, in the clean condition, no other objects were nearby the target object.
    
    \item \textbf{Ambiguity}: Three types of ambiguity, which  could challenge the precise referencing of target object, were explored, i.e., \textit{structural ambiguity}, \textit{positional ambiguity} and \textit{no ambiguity} (baseline). Structural ambiguity refers to ambiguity caused by a structural, complex surface or the fact that the object contains multiple parts. Positional ambiguity refers to  situations where there are multiple similar objects in close proximity to the target object. No ambiguity was free from both structural and positional ambiguities.

\end{itemize}

Based on these three factors, twelve experimental conditions were created (Table \ref{tab:condition}). For each condition, participants completed three trials, selecting a different target object each time\footnote{List of objects can be found in Appendix}. The order of the target objects and conditions were randomized among participants to mitigate order effects. 

Additionally, it's important to mention that we did not compare our technique with other baselines, as the task of physical object referencing is novel and lacks existing baselines for comparison.

\begin{figure*}[ht]
\centering
\includegraphics[width=\textwidth]{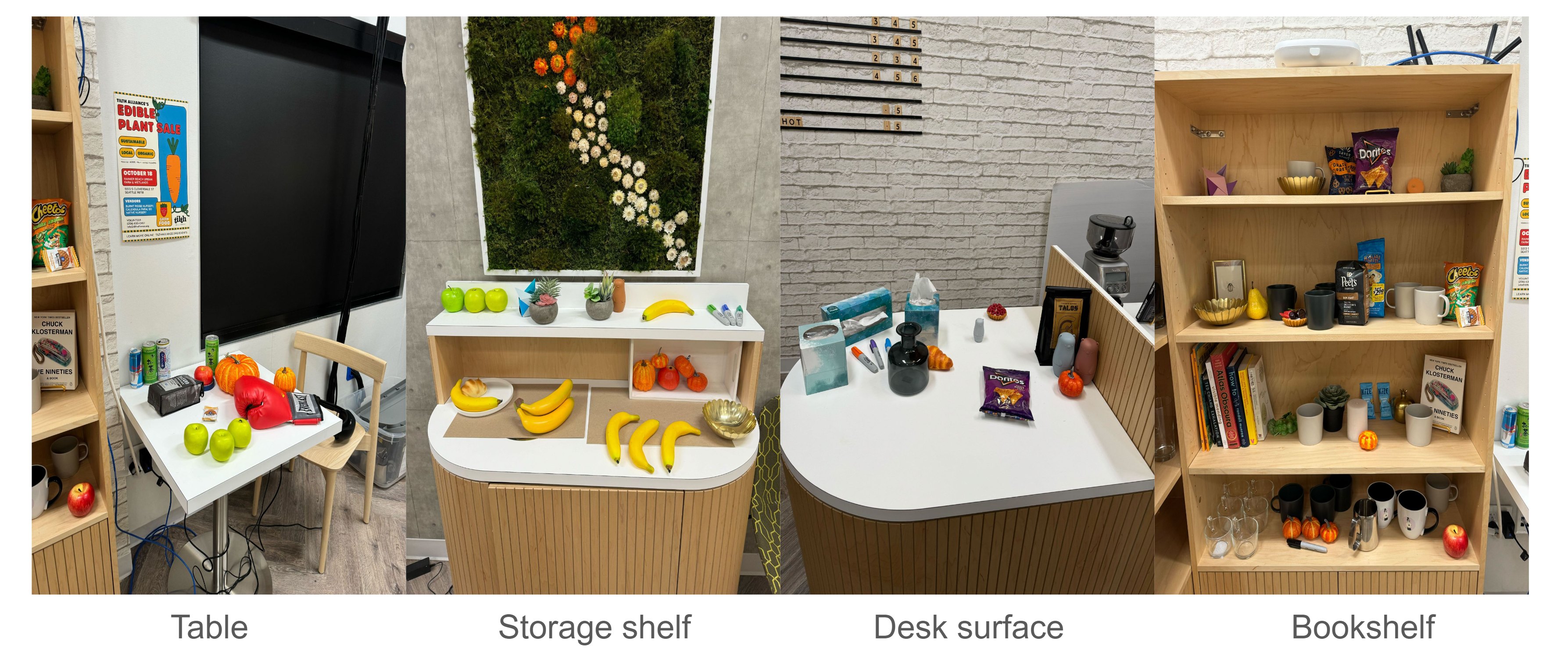}
\caption{Images of the study environment, where participants were asked to select objects using Gazeify Then Voiceify on the table, storage shelf, desk surface, and bookshelf.}
\label{fig:study_environment}
\end{figure*}

\begin{table*}[t]
\centering
\begin{tabular}{|c|c|c|c|}
\hline
\textbf{}         & \textbf{Perceived Object Size} & \textbf{Environmental Clutter} & \textbf{Ambiguity} \\ \hline
\textbf{C1}  & Small     & Cluttered  & Structural   \\ \hline
\textbf{C2}  & Small     & Cluttered  & Positional     \\ \hline
\textbf{C3}  & Small     & Cluttered  & None         \\ \hline
\textbf{C4}  & Small     & Clean    & Structural   \\ \hline
\textbf{C5}  & Small     & Clean    & Positional     \\ \hline
\textbf{C6}  & Small     & Clean    & None         \\ \hline
\textbf{C7}  & Normal    & Cluttered  & Structural   \\ \hline
\textbf{C8}  & Normal    & Cluttered  & Positional     \\ \hline
\textbf{C9}  & Normal    & Cluttered  & None         \\ \hline
\textbf{C10} & Normal    & Clean    & Structural   \\ \hline
\textbf{C11} & Normal    & Clean    & Positional     \\ \hline
\textbf{C12} & Normal    & Clean    & None         \\ \hline
\end{tabular}
\caption{The twelve conditions evaluated during the user study.}
\label{tab:condition}
\end{table*}

\subsection{Study Procedure}

Each study session lasted approximately one hour. Initially, participants signed a consent form and complete a demographic survey. Following this, they underwent eye calibration using the Quest Pro. \zz{We used the Quest Pro's default eye calibration program and asked participants to recalibrate until the calibration error is under $0.5^\circ$}. The experimenters then explain how to use Gazeify Then Voiceify, instructing participants to use the primary index trigger on the controller for gaze selection and the secondary index trigger for voice commands. 

Before starting the trials, participants engaged in two rounds of referencing practice for warm-up. Afterwards, participants were asked to use Gazeify Then Voiceify to select 36 objects in a study room decorated as coffee shop (Figure \ref{fig:study_environment}). The objects were located in four areas in the room (i.e., on a bookshelf, desk surface, kitchen top and storage shelf). During each trial, participants were instructed to initially use their gaze to select the target object and then issue a voice command if they believed the object description was incorrect. For the sake of time, they were permitted a maximum of two rounds of conversation before proceeding to the next trial, regardless of the outcome of their selection.

Once the trials were completed, participants completed a post-study questionnaire and participated in a semi-structured interview with the experimenters. The questionnaire was based on the System Usability Scale~\cite{Bangor2008AnEE}, the NASA TLX~\cite{Hart2006NasaTaskLI}, and had additional questions specific to the Gazeify Then Voiceify user experience. The interview primarily explored the strengths and limitations of the technique, gathered user suggestions, and discussed potential use cases for Gazeify Then Voiceify.

\subsection{Metrics}

We evaluated the following performance metrics to better understand Gazeify Then Voiceify:

\begin{itemize}
    \item \textbf{Gaze referencing accuracy}: The rate at which participant's gaze-based physical object referencing was accurate.
    \item \textbf{Voice description accuracy}: The rate at which participants were able to determine the accuracy of gaze referencing through voice descriptions. This accuracy is important because users can only initiate voice disambiguation when they identify an error in the selection.
    \item \textbf{Voice disambiguation accuracy}: The rate at which participants were able to  use the voice-based disambiguation to correct gaze referencing errors.
\end{itemize}

Regarding the processing time for each stage, the average durations were as follows: gaze-based segmentation required 0.72 seconds using EfficientSAM, voice generation took 3.61 seconds, comprehensive segmentation and filtering lasted 8.68 seconds, and the VLM-based mask update was completed in 5.25 seconds. It's important to note that we do not prioritize trial completion time as a key metric, considering that both segmentation and VLM integration inherently introduced latency into the process.

\subsection{Results}
This section presents the outcomes of gaze referencing, voice description, and disambiguation accuracy, along with common errors identified in each phase. Additionally, it includes user evaluations and feedback regarding the usability, user experience, and potential use cases of this technique.

\subsubsection{Gaze Referencing Accuracy and Common Errors}
\label{sec:factor_effect_gaze_referencing}

\begin{figure*}[ht]
\centering
\includegraphics[width=\textwidth]{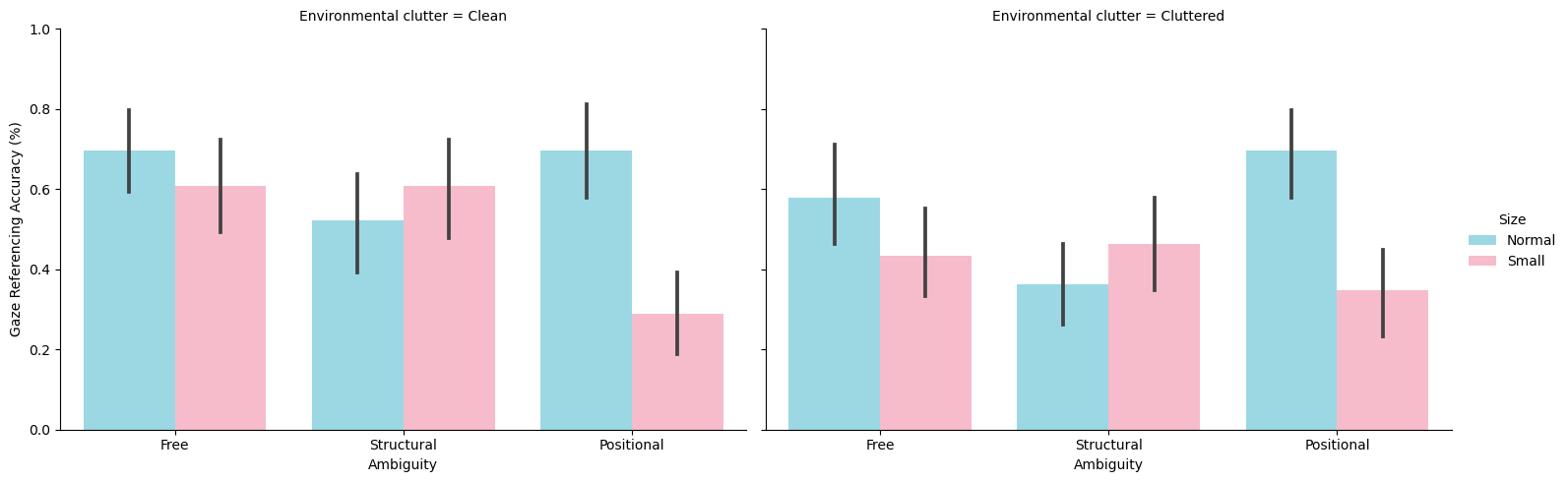}
\caption{Gaze referencing accuracy when different amounts of environmental clutter, perceived object sizes, and ambiguities are present.}
\label{fig:conditional_gaze_accuracy}
\end{figure*}

During the initial phase of gaze-based physical object referencing, participants generated an accurate mask for the target object with an average success rate of 53\% (SD = 13\%). An accurate mask was defined as "an object overlay that accurately delineates the object boundary and covers at least 90\% of the object's 2D area in the captured image frame". A repeated-measures ANOVA was conducted for gaze referencing accuracy and found a significant main effect of accuracy for perceived object size ($F(1, 22)=15.60$, $p<0.001$, $\eta_p^2$=0.415), environmental clutter ($F(1, 22)=4.79$, $p<0.05$, $\eta_p^2$=0.179) and ambiguity ($F(2, 44)=4.51$, $p<0.05$, $\eta_p^2$=0.170; Figure \ref{fig:conditional_gaze_accuracy}). The post-hoc analysis revealed that participants demonstrated higher accuracy in selecting objects of normal size compared to smaller ones. Additionally, as anticipated, accuracy was higher in clean environments, where there was no occlusion or adjacency, as opposed to cluttered environments. Lastly, accuracy was higher in the no ambiguity condition than the positional and structural ambiguity conditions.

We also found an interaction effect between perceived size and ambiguity ($F(2, 44)=22.76$, $p<0.001$, $\eta_p^2$=0.508), and between environmental clutter and ambiguity ($F(2, 44)=4.23$, $p<0.05$, $\eta_p^2$=0.161). To delve deeper into the interaction effects, we then conducted a linear mixed-effects model analysis, treating structural ambiguity, the no ambiguity condition, and positional ambiguity as distinct fixed effects, alongside perceived object size and clutter. Participant ID was incorporated as a random effect. Notably, the analysis revealed that the effect of perceived size was contingent upon the existence of structural ambiguity ($Coef=0.34$, $SD=0.06$, $z=5.11$, $p<0.05$). Interestingly, the post-hoc analysis showed that participants tended to achieve a higher gaze referencing accuracy for small objects than for those with normal size when the object had a complex structure and surface. Upon examining the log data, it was determined that the segmentation model tended to identify noticeable boundaries within parts or distinct sub-regions of large objects. Consequently, the model often returned the mask of the specific sub-region where the gaze point was located. In contrast, the intricate structure of small objects generally appeared indistinguishable to the segmentation model, leading it to segment the entire object more frequently.

We also found the effect of environmental clutter was contingent upon the structural ambiguity ($Coef=-0.45$, $SD=0.18$, $z=-2.27$, $p<0.05$) and no ambiguity ($Coef=-0.26$, $SD=0.24$, $z=-2.17$, $p<0.05$). The post-hoc analysis revealed that in cluttered environments, participants exhibited worse performance when selecting objects in the no ambiguity and structurally ambiguous situations compared to when the environment was clean. However, no interaction effect was observed between environmental clutter and positional ambiguity. This may be due to the fact that, in our study setting, objects with positional ambiguity were also positioned relatively close to other similar objects.

\begin{figure*}[ht]
\centering
\includegraphics[width=\textwidth]{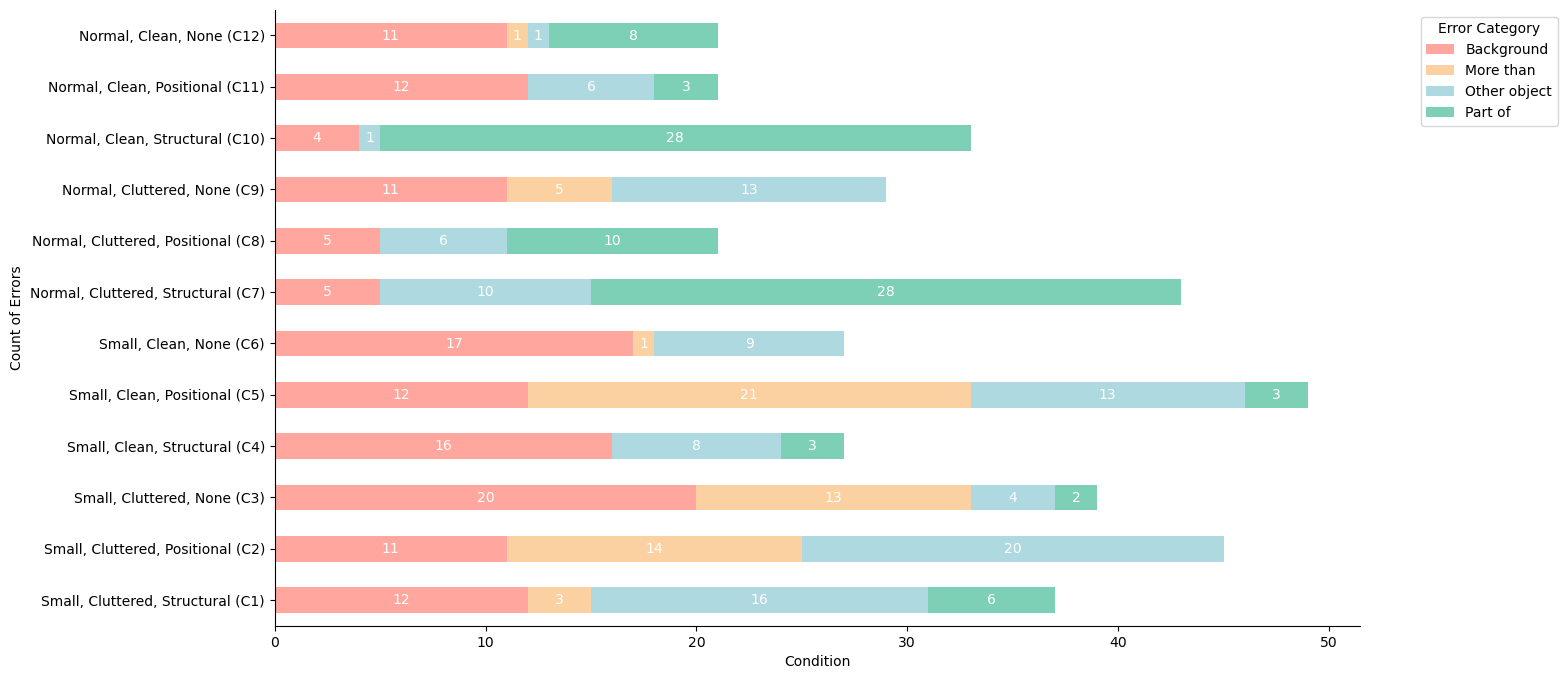}
\caption{Distribution of gaze referencing error types by trial condition.}
\label{fig:gaze_error_distribution}
\end{figure*}

To better understand the low gaze referencing error results, we analyzed the log data to identify the types of gaze referencing errors that were being made. \zz{Note that the study included a total of 828 selection trials.} Overall, we identified four types of gaze referencing errors (Figure \ref{fig:gaze_error_distribution}):

\begin{itemize}
     \item \textbf{Part of errors}: The mask covered only a portion of the target object, rather than its entirety (N = 91).
    \item \textbf{More than errors}: The mask encompassed the intended target object and extraneous objects such as adjacent objects (N = 58).
    \item \textbf{Other object errors}: The gaze selection masked an object other than the target object (N = 107).
    \item \textbf{Background errors}: The gaze selection masked non-target, background elements such as the desktop, surface, or wall (N = 136).   
\end{itemize}


When looking into each type of error, gaze referencing often selects only a portion of a normal-sized object with a complex structure and surface (i.e., C7 and C10), whereas this type of error was less frequent with smaller objects (Figure \ref{fig:gaze_error_distribution}). This observation aligns with the phenomena discussed in Section \ref{sec:factor_effect_gaze_referencing}. Additionally, the "background" error was more prevalent with smaller objects (i.e., C1-6) than with normal-sized ones (i.e., C7-12). Furthermore, from C2, C3, and C5, it is evident that the "more-than" error is more likely to occur when the object is small and situated in a cluttered environment or surrounded by multiple similar objects.

\subsubsection{Voice Description Accuracy and Common Errors}
To assess whether participants could accurately identify if masks were correct based on the system's vocal feedback during the gaze referencing stage, we computed the precision and recall of the voice description accuracy during thi stage. We found a precision of 0.86, a recall of 0.82, an F1 score of 0.84. This suggests that, in the majority of instances, participants could accurately identify if the masks  generated from their gaze referencing were correct using only the voice descriptions provided by the VLM. 

It is important to note that a ground truth bounding box might not be generated or could be incorrectly excluded, leaving the Object Localization model without a way to identify the target object's mask. In such cases, the VLM described the probable target object using the contextual image and user command as outlined in Section \ref{sec:object_localization}. Thus, we also assessed voice description accuracy for the voice disambiguation stage. In the 110 trials where the ground truth bounding box was absent, participants perceived that the system accurately identified and described the target object in 92 (i.e., 83.6\%) of these cases. This indicates that the VLM was still able to correctly interpret the user's command and locate the object, despite failures in the object detection and filtering algorithms. For the 390 rounds of conversation\footnote{Note that participants can make at most two rounds of disambiguation for each trial} where participants provided a command, participants made 25 false positive (i.e., 18 first round, 7 second round) and 32 false negative (i.e., 23 first round, 9 second round) identifications based on the voice description. The precision was 0.89, the recall was 0.86, and the F1 score was 0.88.

\begin{figure*}[ht]
\centering
\includegraphics[width=\textwidth]{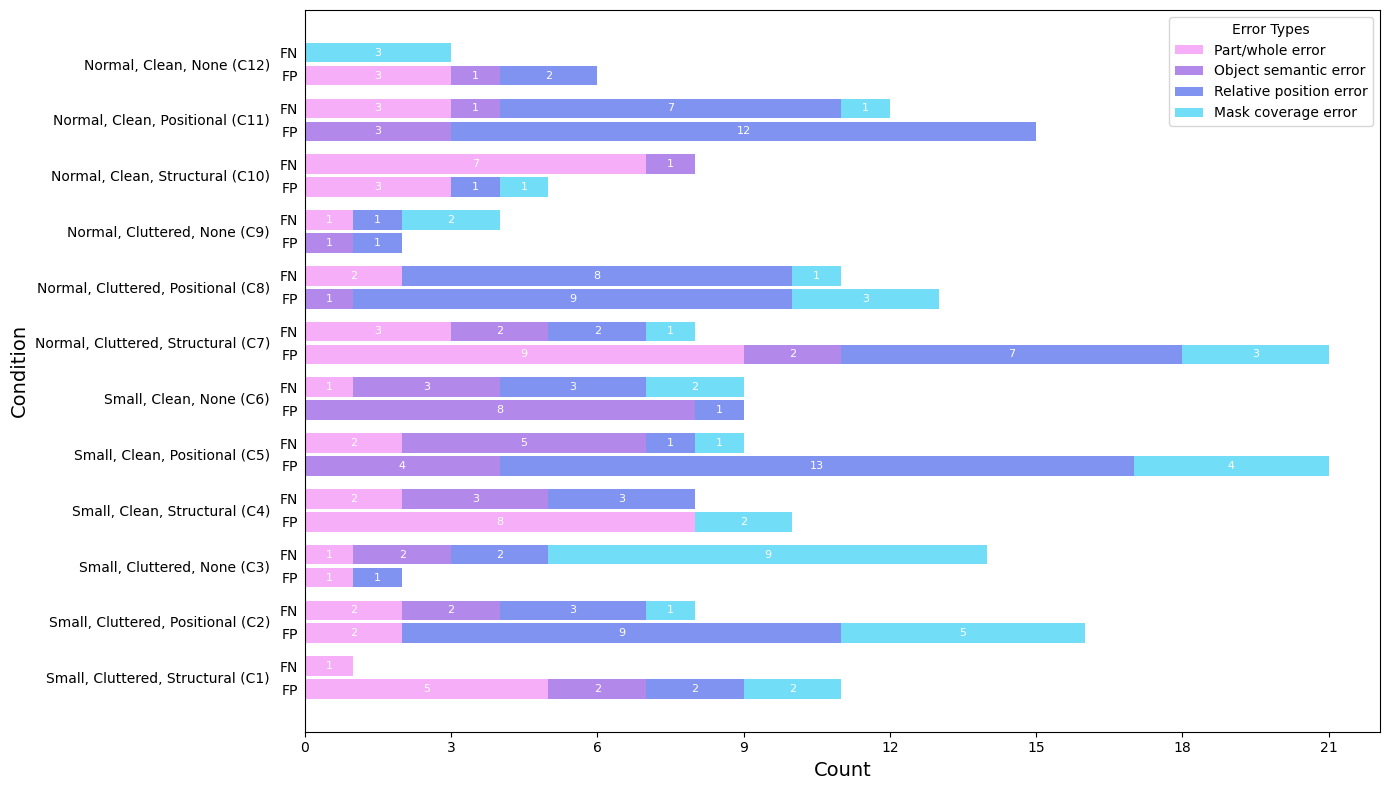}
\caption{The distribution of False Positive (FP) and False (FN) voice description errors across the trial conditions during the gaze referencing and voice disambiguation phases. Notably, instances from the disambiguation phase lacking a ground truth bounding box are excluded from this figure.}
\label{fig:fp_fn_voice_dist}
\end{figure*}

From the log data of the dialog and selection history, we identified four types of voice description errors (Figure \ref{fig:fp_fn_voice_dist}):

\begin{itemize}
    \item \textbf{Part-whole errors}: The description ambiguously represented the hierarchical level of the target object indicated by the mask (e.g., if the mask only displayed a logo on a snack bag, the system incorrectly described the selection as 'a snack bag'; N = 56).
    \item \textbf{Relative position errors}: The description ambiguously specified the identity of the masked object by referencing its position relative to another object (e.g., the phrase "the pumpkin in front of the black cup" was used despite there being three pumpkins located in front of the cup; N = 88).
    \item \textbf{Object semantic errors}: The description ambiguously conveyed the semantic properties and identity of the masked object (e.g., "orange rounded candle" as an "orange cap"; N = 41).
    \item \textbf{Mask coverage errors}: The description inaccurately represented the objects obscured by the mask, either omitting an object that was masked or incorrectly including a nearby object that was not masked (e.g., "a plant and bear toy" even though only the bear toy was masked; N = 41).
\end{itemize}

Part-whole errors frequently occurred when the object had a complex structure or surface, as seen in conditions C1, C4, C7, and C10 (Figure \ref{fig:fp_fn_voice_dist}). Additionally, relative position errors commonly arose when multiple similar objects were positioned closely together, as noted in conditions C2, C5, C8, and C11. This suggests that VLM may struggle to express the position of the selected object distinctly from similar ones using clear spatial relationships. Furthermore, mask coverage and object semantic errors were more likely to occur when participants attempted to select smaller-sized objects compared to normal-sized ones (i.e., C1-6 versus C7-12). This indicates that VLM was less effective at accurately describing smaller mask patches compared to larger ones.


\subsubsection{Voice Disambiguation Phase Results}

Out of 409 trials that included a voice disambiguation phase, participants successfully used voice commands to generate an accurate mask of the target object in 239 of these trials (58.4\%). This success rate was comprised of 177 cases (43.3\%) where participants corrected genuine gaze referencing errors and 62 cases (15.2\%) where they recovered from false negative voice descriptions. On average, each disambiguation conversation consisted of 1.38 rounds, with voice commands averaging 7.82 words in length. During the 239 trials with successful voice disambiguation, the correct mask was achieved using a single voice command in 208 cases (87.0\%), but required two rounds of conversation in 31 (13\%) cases. For the unsuccessful trials, participants were unable to resolve 151 gaze referencing errors (36.9\%) within two rounds of conversation, and they failed to recover from 19 cases (4.65\%) where false negative voice descriptions occurred during the gaze referencing phase.

As the trials where a voice disambiguation phase were necessary were not balanced across trial conditions, we conducted a linear mixed-effect model analysis to better understand how perceived object size, environmental clutter, and ambiguity affected voice disambiguation accuracy. We considered structural ambiguity, positional ambiguity, and no ambiguity as distinct fixed effects, along with perceived object size and environmental clutter. Participant ID was a random effect. Our findings indicated that perceived object size significantly impacted voice disambiguation accuracy ($Coef=-0.27$, $SD=0.17$, $z=-1.59$, $p<0.05$), suggesting that participants are more likely to struggle with the disambiguation of smaller objects that normal sized ones. We did not observe significant effects from the other factors, nor did we detect any interaction effects.

\begin{figure*}[t]
\centering
\includegraphics[width=\textwidth]{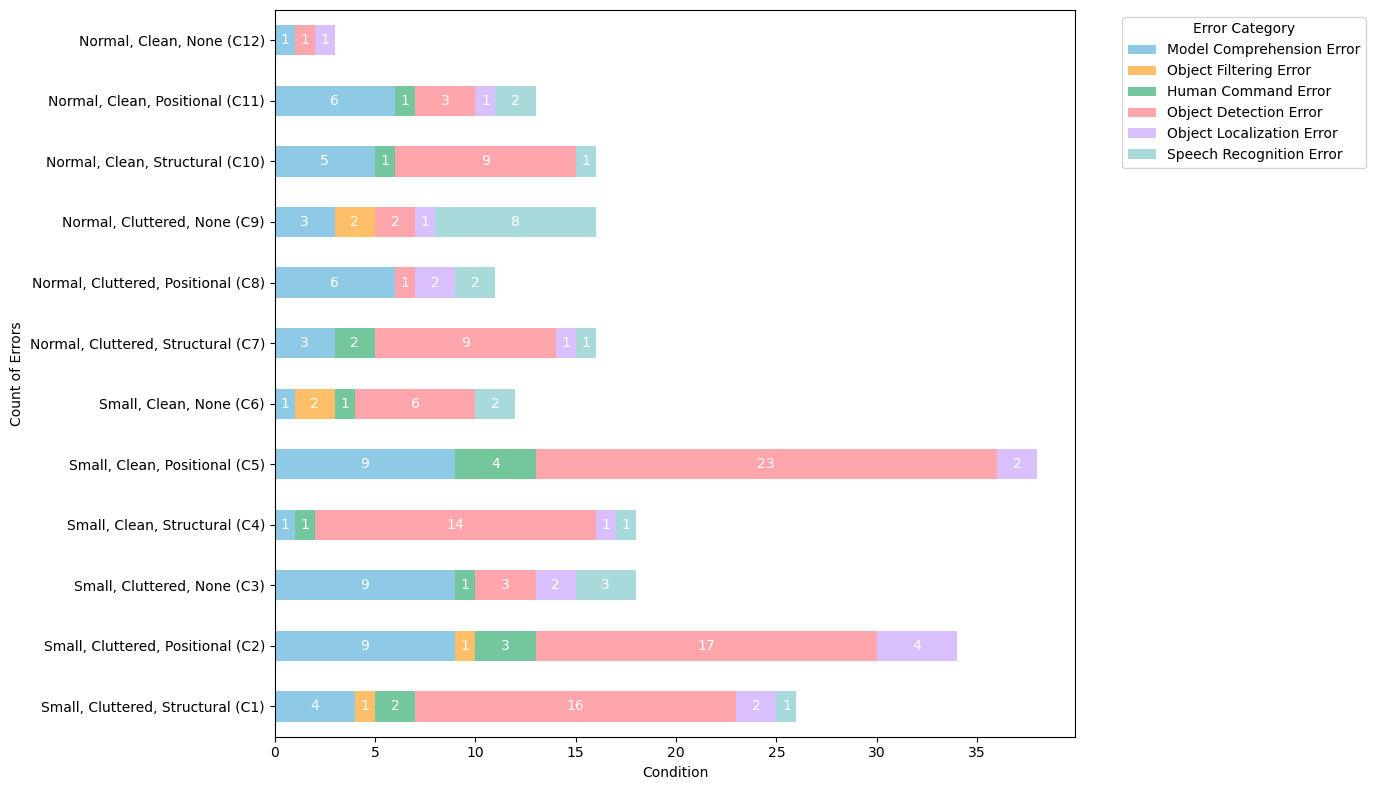}
\caption{The distribution of voice disambiguation error types across different trial conditions.}
\label{fig:voice_ambigutation_error_dist}
\end{figure*}

A review of the failed voice disambiguation phases resulted in the identification of six types of errors (Figure \ref{fig:voice_ambigutation_error_dist}):

\begin{itemize}
    \item \textbf{Object detection errors}: The target object was not detected during the detection phase (N = 104).
    \item \textbf{Object filtering errors}: The target object was detected but incorrectly filtered out (N = 6).
    \item \textbf{Human command errors}: the participant provided an uninformative command, such as "try again" or "it is not what I want to select" (N = 16).
    \item \textbf{Speech recognition errors}: The speech recognition inaccurately translated the participant's command (N = 21).
    \item \textbf{Model comprehension errors}: The VLM misinterpreted the participant's command, leading to errors such as incorrect pronoun resolution or the misidentification of the target object, reference object, or their spatial relationship (N = 57).
    \item \textbf{Object localization errors}: The VLM incorrectly identified a different object as the target mentioned in the participant's command during the localization phase (e.g., the VLM identified that the participant wanted to select the leftmost marker amongst the three but it returned the middle one instead; N = 17).
\end{itemize}

Object detection errors were common when the target object was small, particularly in comparisons between C1-6 and C7-12. This issue was exacerbated when there was positional or structural ambiguity when selecting small objects, as observed in C1, C2, C4, and C5. These findings indicate that our detection method based on leading segmentation and object detection models could struggle to accurately segment small adjacent or complex objects. The second most common type of error were model comprehension errors, which arose from positional ambiguity and cluttered environments, as evidenced in C1, C2, C3, C5, C8, and C11. Many model comprehension errors also stemmed from unclear commands given by participants. For instance, users would say ``the right pumpkin'' instead of specifying ``the rightmost pumpkin'' when multiple pumpkins were stacked together. In other cases, the model incorrectly identified the target object, spatial relationships, or reference objects. For example, the VLM misidentified ``carrot'' as the target object when the participant meant ``carrot poster'', and misinterpreted the spatial relationship in the command ``the middle marker behind the vase'' as ``middle'', even though ``behind'' referred to the spatial relationship with the reference object.



\paragraph{Qualitative analysis of user command}

We also carried out a qualitative analysis of user commands during the voice disambiguation phase. We observed that the level of detail in participants' commands varied based on environmental factors and the selection errors they identified from the descriptions. When the voice description suggested that an incorrect object or background was selected, participants typically used simple commands targeting the desired object, such as "select the apple" or "the white cup." When faced with multiple similar objects in close proximity, participants more frequently used spatial and ordinal terms to specify their choice. Additionally, they often combined the use of pronouns for the target or reference object with spatial and ordinal descriptors, using phrases like "the one on the left," "the rightmost pumpkin among them," or "the purple marker behind it."

\subsubsection{Usability, User Experience, and Potential Use Cases}

\begin{figure*}[ht]
\centering
\includegraphics[width=\textwidth]{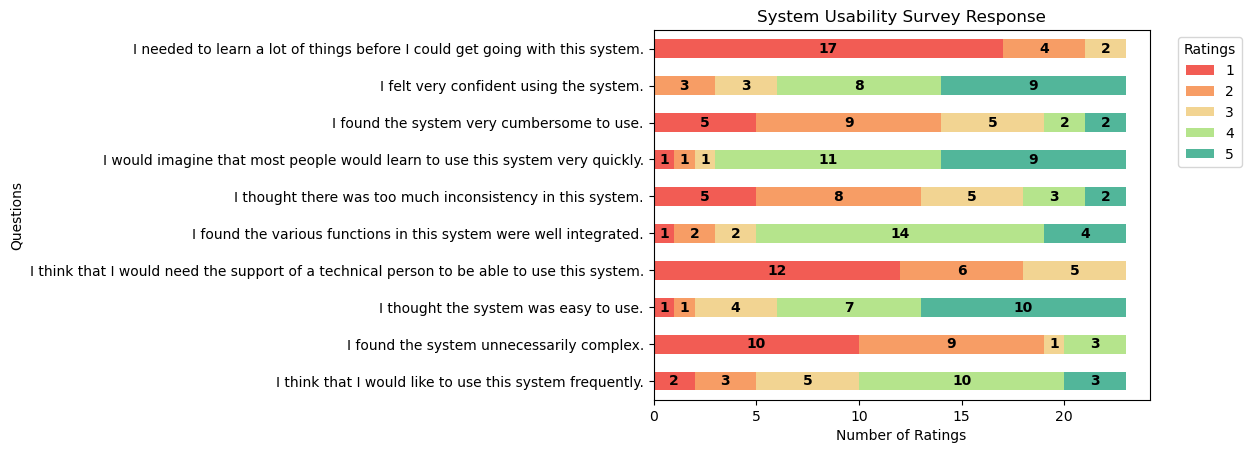}
\caption{The results of the System Usability Scale, including the mean and standard deviation for the ratings of each question. Odd numbered questions are reverse scored.}
\label{fig:sus_survey_result}
\end{figure*}

\begin{figure*}[ht]
\centering
\includegraphics[width=\textwidth]{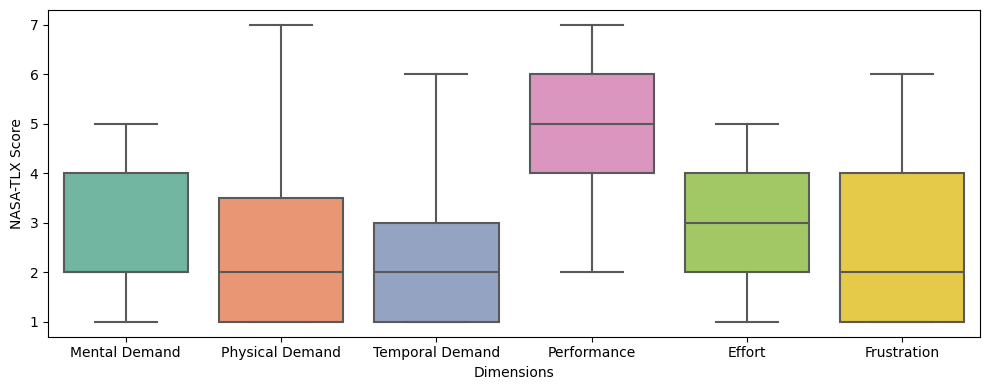}
\caption{The results of the NASA-TLX survey.}
\label{fig:nasa_tlx_survey}
\end{figure*}

\begin{figure*}[ht]
\centering
\includegraphics[width=\textwidth]{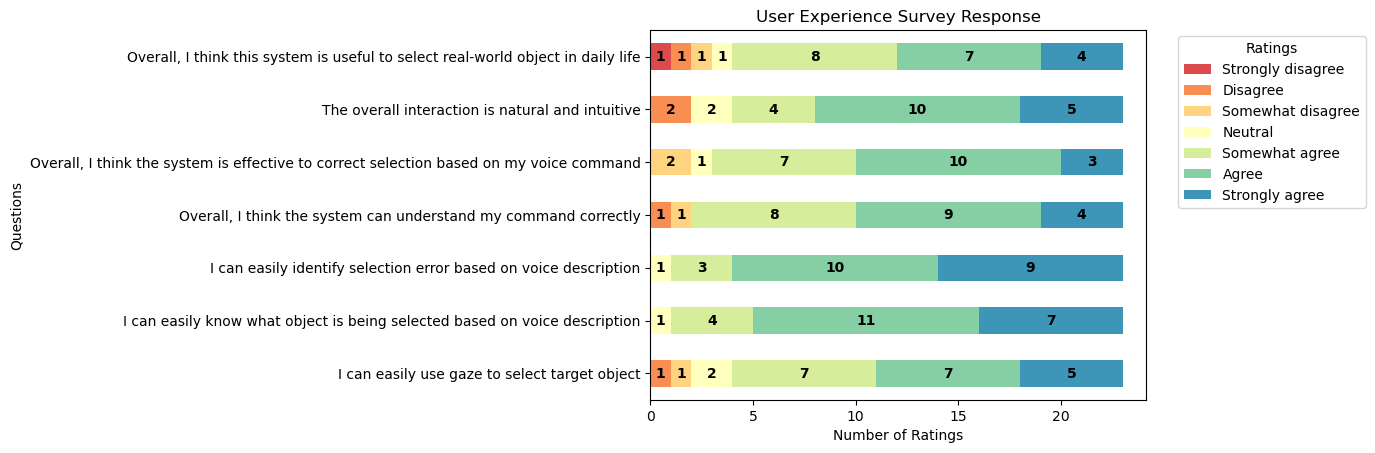}
\caption{The results of the user experience and effectiveness questions.}
\label{fig:user_experience_survey}
\end{figure*}

Figure \ref{fig:sus_survey_result} shows the participants' ratings of System Usability Scale (SUS) survey. The average SUS score for the system was 73.70 (SD=16.56), it indicates the system has a good usability. Additionally, the NASA-TLX survey results indicated that participants experienced low levels of mental and physical demand, as well as reduced frustration, difficulty, and temporal pressure (Figure \ref{fig:nasa_tlx_survey}). Lastly, participants found the gaze referencing and voice disambiguation technique to be useful, natural, and effective for selecting physical objects in daily life using smart glasses (Figure \ref{fig:user_experience_survey}).

Participants recognized many strengths of the Gazeify Then Voiceify technique. First, despite the efficiency of gaze-based selection, participants valued the ability to use voice to refine their selections. P1 noted, "\textit{it would cause eye fatigue if I always used gaze to select, the voice interaction makes it less effortful since I can simply look at the object casually and use voice to clarify if the selection is correct}". Another frequently mentioned advantage was the clear and accurate description of the selection results. Participants were impressed by the system's ability to correctly recognize the selected object and clarify its properties and position within the immediate environment. For example, P19 remarked, "\textit{the accuracy of the voice description surprises me a lot, it describes the right color, position, and identity of the object I selected and it is easy to distinguish it from the surrounding items}". Additionally, participants found that they could easily learn the most effective ways to verbally instruct the system after a few interactions with the Voiceify component. For example, P14 said ``\textit{after a few rounds I realized articulating the color or position relative to nearby object is very useful to help the system get to the target}''. Lastly, participants appreciated the system's efficiency in resolving ambiguities by providing one piece of feedback in most cases. As P2 expressed, "\textit{the system is able to correct itself with a simple command and I don't need to go back and forth}".

Participants also identified several challenges and limitations. First, participants noticed that the voice disambiguation sometimes broke down when there were several similar objects close to each other or there were several small objects, which made them frustrated. Some participants additionally complained that in some cases the voice description was not specific, for example, it could ``\textit{use ambiguous terms such as `next to something' or `rectangular object' which is unclear to distinguish which object it refers to}'' (P7). At the same time, some participants (P10, P16, P21) found it was unnecessary to always provide a verbose description, especially when the selected object was very obvious and the environment was simple. Participants also felt it was frustrating when the system kept making mistakes on an obvious object selection, like a standalone poster and snack bag. An expert participant (P14) also believed using this system requires some basic knowledge about how the segmentation model and object detection models work to understand why the system would make an obviously wrong selection (e.g. only select a logo of a bag). Other participants (P3, P12, P23) found it was exhausting to perform several precise gaze references, as P3 said, ``\textit{human gaze could saccade unintentionally, I have to really make sure I'm holding my gaze when selecting the object and it is pretty cognitive stressful after many attempts}''. Lastly, participants also reported the response latency was lengthy for the first round of voice disambiguation, which led them to become impatient after several trials.

Several suggestions were made to enhance the system's user experience. First, many suggested that it would be beneficial if users could see a visual representation of their selection on a smartphone or smartwatch, although this might introduce new assumptions about device availability. Additionally, participants suggested that for clearly identifiable objects, the system should utilize commonsense knowledge to determine the target object despite any gaze drift. For instance, P15 remarked, "\textit{for the marker cases, a smart system should easily recognize that the entire marker should be the target, not just the cap}". Furthermore, P18 suggested that the system should adapt to individual user behaviors and provide personalized voice feedback or automatically correct consistent gaze errors. P3 proposed enabling users to employ simple gestures, such as swiping and zooming, to swiftly resolve common positional and structural ambiguities. Lastly, many participants expressed a desire for quicker system feedback during the disambiguation phase or for cues that indicate the potential waiting time.

Our participants suggested several potential use cases the Gazeify Then Voiceify technique:

\begin{itemize}
    \item \textbf{Collaborative Referencing}: Participants believed that the technique could enhance their ability to communicate references to objects within their environment during collaborative activities such as cooking, meetings, and gaming. For instance, P1 found the technique beneficial for quickly sharing digital copies of physical objects or content, such as drawings on a whiteboard or printouts, during remote meetings. Additionally, P5 described a scenario where he would use the technology while cooking with his wife, allowing him to highlight specific ingredients or materials he would use so his wife could concentrate on other tasks. Lastly, P11 saw the potential for AR gaming with her son (e.g., \textit{"My son sometimes gets frustrated with me when I'm not sure which object he is referring to while playing a family AR game. Now, he can use gaze to quickly show what he means, and we can collaborate more efficiently."}).

    \item \textbf{Shopping}: Participants also believed that the technique could enhance their daily shopping experiences. For instance, P2 mentioned that she could swiftly select items on cluttered shelves and check their online prices to ensure she gets the best deal. Additionally, P19 noted that he could use this technique to obtain information about items that are out of reach, such as nutritional details, without having to physically retrieve them.

    \item \textbf{Digital Memories or Stickers Creation}: Participants also saw potential for creating digital memories or stickers of physical objects for entertainment or when compiling an everyday repository for future use. For instance, P6 explained, "\textit{My wife can use it to easily create a checklist of clothes she wants to donate and share it with me to pack them up}". Additionally, P17 thought it would be enjoyable to create stickers from his surroundings and send them to friends. Lastly, P20 mentioned that the technique could assist her in saving the locations of items she frequently misplaces, allowing her to simply ask the glasses to locate them when needed.

    \item \textbf{Augmented Interaction with IoT Devices and AI}: Participants also suggested that the technique could be used to direct IoT devices or AI. For instance, P11 envisioned using gaze to select an item for a family smart robot to retrieve. P22 believed that this technique could integrate AI with one's immediate surroundings, e.g., \textit{"I enjoy painting, and I can now guide the AI to intelligently incorporate real-world elements into my artwork"}.
    
\end{itemize}

\section{DISCUSSION}

\subsection{Achieving Referencing Efficiency and Accuracy With Multiple Modalities}

The ideal referencing technique should enable users to select targets efficiently and accurately. However, when using head-mounted devices, these two objectives often conflict. For instance, gaze input offers a rapid selection experience but suffers from low accuracy due to gaze and eye-tracking noise. Conversely, head pointing and ray casting typically ensure more accurate selections, though they generally take longer to complete than gaze selections. Previous research has explored methods of combining gaze with other input (e.g., Pinpointing\cite{kyto2018pinpointing}, Gaze-Hand Alignment\cite{lystbaek2022gaze}) to achieve  efficiency and accuracy through a multi-stage process. With this process, gaze is used for quick, initial selection, followed by the other modalities to resolve any ambiguities. These techniques assume that continuous visual feedback is available to inform the user of the selection progress.

Our research introduces a new technique that enhances the efficiency and accuracy of referencing physical objects with displayless smart glasses. \zz{In contrast to gaze and hand interactions \cite{tutuncu2025handover, zhang2025forcepinch} that rely on visual feedback, we leverage the unique affordances of this hardware, specifically its egocentric camera, unobtrusive form factor to provide AI-driven semantic context and microphone for voice interaction.} Gazeify utilizes gaze-based selection for the quick identification and segmentation of physical objects, while Voiceify integrates a voice channel that allows users to describe and correct selections through free-form commands. This system preserves the speed of gaze selection while enabling natural voice interactions for corrections without a display

Despite the generally positive user experience and high success rates, our study identified several limitations related to the current technology used in our technique. First, some participants felt that the voice descriptions were overly verbose, despite providing detailed information on their selections. This suggests the need for a more adaptive voice description generation method, where the response length can be adjusted based on the complexity of the environment and objects involved. Additionally, the efficiency of gaze referencing is somewhat reduced by the latency in the VLM, segmentation, and object detection model. Moreover, the effectiveness of disambiguation is limited by the capabilities of the commercial models employed, particularly in terms of object detection accuracy and the VLM's ability to comprehend images and commands. We believe these issues could be alleviated through the development and adoption of more efficient end-to-end language-driven models for segmentation and detection.

\subsection{Accessible Object Referencing For People With Reduced Abilities}

Some study participants emphasized the potential benefits of our technique for individuals with limited mobility, such as the elderly. Our technique could enable them to interact remotely with objects that are out of reach. Considering that elderly individuals often face low-vision challenges their gaze input, would, however, be imprecise. In such cases, voice disambiguation would become even more crucial to confirm and verbally correct their selections. This would ensure that  subsequent interactions with IoT devices or commands to mobile robots to engage with selected objects would be accurate. This technique could complement existing IoT interaction methods such as manipulation interfaces \cite{cabrera2021exploration}, voice~\cite{seaborn2021voice} and mid-air gestures~\cite{locken2012user, zaicti2015free, ackad2015wild}, and enable users to interact with commonplace objects in their environment more effectively.

Future research could enhance the physical object referencing system to more effectively cater to users with limited abilities. For instance, integrating haptic feedback into smart glasses could provide additional sensory information about the properties of objects targeted by gaze, such as size, shape, and texture. This would assist users in discerning their selections more clearly. Additionally, incorporating an automatic image zooming function in the background would be beneficial for users who are unable to physically move closer to an object.

\subsection{Design Implications for Everyday Assistance}

\zz{Beyond technical accuracy, our study results and participant feedback highlight how the gaze with voice paradigm transforms smart glasses into active agents for everyday assistance. By moving beyond simple information display, this multimodal interaction offers three key design implications for future ubiquitous computing:}

\zz{\textbf{Augmenting Shared Attention and Collaboration}. While traditional HMDs often isolate users, our participants highlighted the potential for gaze voice interaction to enhance social connection through ``collaborative referencing''. By utilizing gaze to segment an object and voice to share it, users can bridge the gap between their egocentric view and a collaborator's context. Participants envisioned scenarios such as sharing whiteboard content in remote meetings (P1), identifying ingredients while cooking with a spouse (P5), or clarifying targets in family AR games (P11). This implies that future smart glasses should be designed not just as personal information retrieval devices, but as communication tools that externalize the user's visual attention to facilitate shared understanding.}

\zz{\textbf{Extending Physical Reach and Interaction.} The combination of gaze selection and semantic voice description effectively acts as a virtual ``cursor'' for the physical world, allowing users to interact with objects beyond their immediate physical reach. Participants noted the utility of checking prices on high shelves or accessing nutritional information for out-of-reach items without physical retrieval (P19). Furthermore, this paradigm extends to controlling IoT devices, where gaze serves as the pointer and voice as the command, such as instructing a robot to retrieve a specific item (P11) or guiding AI-generated art based on real-world elements (P22). This suggests a design shift where smart glasses serve as a universal remote control for the ambient environment.}

\zz{\textbf{Seamless Digitization of Physical Context.} Finally, the system’s ability to generate semantic masks enables the effortless capture and organization of physical memories. Unlike taking a photo, which captures a whole scene, Gazeify allows for the extraction of specific semantic items—effectively creating ``digital stickers'' of reality. Participants proposed using this for creating donation checklists (P6), saving the location of misplaced items (P20), or creating entertainment content (P17). This implies a future use case where smart glasses function as an ``always-on'' semantic filter, allowing users to catalogue and retrieve knowledge about their physical environment through natural observation and conversation.}
\section{LIMITATION \& FUTURE WORK}

The current version of Gazeify Then Voiceify has several constraints. First, the current system can only select a single object as a reference. In reality, users may want to perform multi-object referencing such as group selection or selecting non-adjacent objects. Second, the current system performs poorly when selecting object at long distances (> 3 meters) because the perceived object size is reduced and the gaze point could be off the target. Also, the segmentation and object detection model have a difficult time detecting small objects precisely. Third, we noticed sometimes a hallucination issue happened with the VLM where a user selected a wrong candidate object bounding box while describing it as the same user command requests. Lastly, though Gazeify can respond in nearly real-time, the Voiceify could introduce latency that would lessen one's user experience. \zz{To reduce latency, future systems can use lightweight on-device models for faster inference.}

In addition, the current system has some implementation limitations. \zz{We used the Quest Pro as a proxy for displayless smart glasses. However, we acknowledge the gap between this HMD proxy and the intended device class. The Quest Pro’s bulkier form factor and video passthrough cannot fully replicate the optical transparency, lightweight comfort, or social acceptance of actual smart glasses.} The low-resolution display in pass-through mode also causes eye fatigue and could increase the likelihood of gaze drift and lessen usability. The EfficientSAM and REDETR background models were also running on a laptop connected to the headset with a cable. Therefore, the current implementation would not support wireless object referencing.

Regarding the study design, there are several potential threats to the validity of our findings. First, we fixed the distance of the object referencing to ensure consistency between sessions, however, it may not have reflected the accuracy of the system in contexts where users could perform referencing from any distance. Although we tested the system for various complex situations, the study result could still depend on the position, type and characteristics of the pre-defined target objects or the study room environment. For example, the segmentation and object detection model could be favorable to certain types of objects so that they could have a better chance of being detected. In addition, as we used fixed hyperparameters such as the image cropped size and confidence threshold for object localization, this could have introduce bias into the study results. Besides, we did not establish a baseline condition to compare "Gazeify Then Voiceify" against, such as head pointing or gestures, because our focus was on evaluating the performance of our technique across various complexities within the constrained time frame of the study sessions. Consequently, the advantages and disadvantages of this technique compared to more traditional methods remains unclear. \zz{Lastly, regarding the study protocol, participants were not provided with specific voice command formats during the warm-up sessions to optimize the performance of Voiceify. While this decision was intended to test the naturalness of the system's free-form interaction, the lack of sample commands may have contributed to the frustration participants reported when they failed to resolve selection errors within the allotted two rounds of voice corrections.}

In the future, we want to further run a technical evaluation of the effectiveness of our pipeline with different kinds of object detection and segmentation models. We want to fine-tune the EfficientSAM model with user gaze data to increase the segmentation accuracy when noisy gaze input is provided. Additionally, we would like to develop an adaptive voice disambiguation method that tailors voice descriptions to include only essential details, streamlining the interaction process. We also intend to integrate a layer of common sense reasoning into Gazeify, to enable it to autonomously correct clear selection errors without requiring additional human intervention. Finally, we want to explore the possibility of running segmentation and object detection models directly on the device, facilitating the use of the system in a wireless mode and increasing its versatility and accessibility.
\section{CONCLUSION}

In this paper, we introduce Gazeify Then Voiceify, a novel method for referencing and disambiguating physical objects using displayless smart glasses. Our user study showed that participants successfully completed gaze selection in 53\% of the trials and verbally corrected 58\% of the remaining errors across various conditions. Furthermore, the system was perceived as likable, useful, and user-friendly. The findings suggest that future smart glasses could enable gaze-based selection for objects that are normal-sized and simply structured in clean environments. However, for more complex scenarios, an intelligent assistance or disambiguation technique, such as voice-based interaction, may be necessary.

\section*{GenAI Usage Disclosure}

We used Generative AI to refine the manuscript's writing and clarity. Additionally, a Vision Language Model (VLM) serves as a core interactive component in our "Voiceify" system, responsible for generating object descriptions and processing voice commands to correct selections. The conceptual framework, system architecture, and experimental design are the original work of the authors.

\bibliographystyle{ACM-Reference-Format}
\bibliography{main}


\begin{thebibliography}{116}


\ifx \showCODEN    \undefined \def \showCODEN     #1{\unskip}     \fi
\ifx \showDOI      \undefined \def \showDOI       #1{#1}\fi
\ifx \showISBNx    \undefined \def \showISBNx     #1{\unskip}     \fi
\ifx \showISBNxiii \undefined \def \showISBNxiii  #1{\unskip}     \fi
\ifx \showISSN     \undefined \def \showISSN      #1{\unskip}     \fi
\ifx \showLCCN     \undefined \def \showLCCN      #1{\unskip}     \fi
\ifx \shownote     \undefined \def \shownote      #1{#1}          \fi
\ifx \showarticletitle \undefined \def \showarticletitle #1{#1}   \fi
\ifx \showURL      \undefined \def \showURL       {\relax}        \fi
\providecommand\bibfield[2]{#2}
\providecommand\bibinfo[2]{#2}
\providecommand\natexlab[1]{#1}
\providecommand\showeprint[2][]{arXiv:#2}

\bibitem[Ackad et~al\mbox{.}(2015)]%
        {ackad2015wild}
\bibfield{author}{\bibinfo{person}{Christopher Ackad}, \bibinfo{person}{Andrew Clayphan}, \bibinfo{person}{Martin Tomitsch}, {and} \bibinfo{person}{Judy Kay}.} \bibinfo{year}{2015}\natexlab{}.
\newblock \showarticletitle{An in-the-wild study of learning mid-air gestures to browse hierarchical information at a large interactive public display}. In \bibinfo{booktitle}{\emph{Proceedings of the 2015 ACM International Joint Conference on Pervasive and Ubiquitous Computing}}. \bibinfo{pages}{1227--1238}.
\newblock


\bibitem[Angelo et~al\mbox{.}(1991)]%
        {Angelo1991ComparingTH}
\bibfield{author}{\bibinfo{person}{Jennifer Angelo}, \bibinfo{person}{Curtis~L. Deterding}, {and} \bibinfo{person}{Jerry Weisman}.} \bibinfo{year}{1991}\natexlab{}.
\newblock \showarticletitle{Comparing three head-pointing systems using a single subject design.}
\newblock \bibinfo{journal}{\emph{Assistive technology : the official journal of RESNA}}  \bibinfo{volume}{3 2} (\bibinfo{year}{1991}), \bibinfo{pages}{43--9}.
\newblock
\urldef\tempurl%
\url{https://api.semanticscholar.org/CorpusID:40600540}
\showURL{%
\tempurl}


\bibitem[Awais et~al\mbox{.}(2023)]%
        {awais2023foundational}
\bibfield{author}{\bibinfo{person}{Muhammad Awais}, \bibinfo{person}{Muzammal Naseer}, \bibinfo{person}{Salman Khan}, \bibinfo{person}{Rao~Muhammad Anwer}, \bibinfo{person}{Hisham Cholakkal}, \bibinfo{person}{Mubarak Shah}, \bibinfo{person}{Ming-Hsuan Yang}, {and} \bibinfo{person}{Fahad~Shahbaz Khan}.} \bibinfo{year}{2023}\natexlab{}.
\newblock \showarticletitle{Foundational models defining a new era in vision: A survey and outlook}.
\newblock \bibinfo{journal}{\emph{arXiv preprint arXiv:2307.13721}} (\bibinfo{year}{2023}).
\newblock


\bibitem[B{\^a}ce et~al\mbox{.}(2016)]%
        {bace2016ubigaze}
\bibfield{author}{\bibinfo{person}{Mihai B{\^a}ce}, \bibinfo{person}{Teemu Lepp{\"a}nen}, \bibinfo{person}{David~Gil De~Gomez}, {and} \bibinfo{person}{Argenis~Ramirez Gomez}.} \bibinfo{year}{2016}\natexlab{}.
\newblock \showarticletitle{ubiGaze: ubiquitous augmented reality messaging using gaze gestures}.
\newblock In \bibinfo{booktitle}{\emph{SIGGRAPH ASIA 2016 Mobile Graphics and Interactive Applications}}. \bibinfo{pages}{1--5}.
\newblock


\bibitem[Bangor et~al\mbox{.}(2008)]%
        {Bangor2008AnEE}
\bibfield{author}{\bibinfo{person}{Aaron Bangor}, \bibinfo{person}{Philip~T. Kortum}, {and} \bibinfo{person}{James~T. Miller}.} \bibinfo{year}{2008}\natexlab{}.
\newblock \showarticletitle{An Empirical Evaluation of the System Usability Scale}.
\newblock \bibinfo{journal}{\emph{International Journal of Human–Computer Interaction}}  \bibinfo{volume}{24} (\bibinfo{year}{2008}), \bibinfo{pages}{574 -- 594}.
\newblock
\urldef\tempurl%
\url{https://api.semanticscholar.org/CorpusID:29843973}
\showURL{%
\tempurl}


\bibitem[Bansal et~al\mbox{.}(2018)]%
        {bansal2018zero}
\bibfield{author}{\bibinfo{person}{Ankan Bansal}, \bibinfo{person}{Karan Sikka}, \bibinfo{person}{Gaurav Sharma}, \bibinfo{person}{Rama Chellappa}, {and} \bibinfo{person}{Ajay Divakaran}.} \bibinfo{year}{2018}\natexlab{}.
\newblock \showarticletitle{Zero-shot object detection}. In \bibinfo{booktitle}{\emph{Proceedings of the European conference on computer vision (ECCV)}}. \bibinfo{pages}{384--400}.
\newblock


\bibitem[Beckmann et~al\mbox{.}(2023)]%
        {Beckmann2023SAMMG}
\bibfield{author}{\bibinfo{person}{Daniel Beckmann}, \bibinfo{person}{Jacqueline Kockwelp}, \bibinfo{person}{J{\"o}rg Gromoll}, \bibinfo{person}{Friedemann Kiefer}, {and} \bibinfo{person}{Benjamin Risse}.} \bibinfo{year}{2023}\natexlab{}.
\newblock \showarticletitle{SAM meets Gaze: Passive Eye Tracking for Prompt-based Instance Segmentation}. In \bibinfo{booktitle}{\emph{Gaze Meets ML}}.
\newblock
\urldef\tempurl%
\url{https://api.semanticscholar.org/CorpusID:269648349}
\showURL{%
\tempurl}


\bibitem[Bhowmick et~al\mbox{.}(2021)]%
        {Bhowmick2021UnderstandingGP}
\bibfield{author}{\bibinfo{person}{Shimmila Bhowmick}, \bibinfo{person}{Keyur Sorathia}, {and} \bibinfo{person}{Pratul~Chandra Kalita}.} \bibinfo{year}{2021}\natexlab{}.
\newblock \showarticletitle{Understanding Gesture Performance for Object Selection in VR: Classification and Taxonomy of Gestures in HCI}.
\newblock \bibinfo{journal}{\emph{Proceedings of the 12th Indian Conference on Human-Computer Interaction}} (\bibinfo{year}{2021}).
\newblock
\urldef\tempurl%
\url{https://api.semanticscholar.org/CorpusID:247085066}
\showURL{%
\tempurl}


\bibitem[Blattgerste et~al\mbox{.}(2018)]%
        {blattgerste2018advantages}
\bibfield{author}{\bibinfo{person}{Jonas Blattgerste}, \bibinfo{person}{Patrick Renner}, {and} \bibinfo{person}{Thies Pfeiffer}.} \bibinfo{year}{2018}\natexlab{}.
\newblock \showarticletitle{Advantages of eye-gaze over head-gaze-based selection in virtual and augmented reality under varying field of views}. In \bibinfo{booktitle}{\emph{Proceedings of the Workshop on Communication by Gaze Interaction}}. \bibinfo{pages}{1--9}.
\newblock


\bibitem[Bolt(1980)]%
        {Bolt1980PutthatthereVA}
\bibfield{author}{\bibinfo{person}{Richard~A. Bolt}.} \bibinfo{year}{1980}\natexlab{}.
\newblock \showarticletitle{“Put-that-there”: Voice and gesture at the graphics interface}. In \bibinfo{booktitle}{\emph{International Conference on Computer Graphics and Interactive Techniques}}.
\newblock
\urldef\tempurl%
\url{https://api.semanticscholar.org/CorpusID:15450886}
\showURL{%
\tempurl}


\bibitem[Cabrera et~al\mbox{.}(2021)]%
        {cabrera2021exploration}
\bibfield{author}{\bibinfo{person}{Maria~E Cabrera}, \bibinfo{person}{Tapomayukh Bhattacharjee}, \bibinfo{person}{Kavi Dey}, {and} \bibinfo{person}{Maya Cakmak}.} \bibinfo{year}{2021}\natexlab{}.
\newblock \showarticletitle{An exploration of accessible remote tele-operation for assistive mobile manipulators in the home}. In \bibinfo{booktitle}{\emph{2021 30th IEEE International Conference on Robot \& Human Interactive Communication (RO-MAN)}}. IEEE, \bibinfo{pages}{1202--1209}.
\newblock


\bibitem[Cai et~al\mbox{.}(2025)]%
        {cai2025aiget}
\bibfield{author}{\bibinfo{person}{Runze Cai}, \bibinfo{person}{Nuwan Janaka}, \bibinfo{person}{Hyeongcheol Kim}, \bibinfo{person}{Yang Chen}, \bibinfo{person}{Shengdong Zhao}, \bibinfo{person}{Yun Huang}, {and} \bibinfo{person}{David Hsu}.} \bibinfo{year}{2025}\natexlab{}.
\newblock \showarticletitle{AiGet: Transforming Everyday Moments into Hidden Knowledge Discovery with AI Assistance on Smart Glasses}. In \bibinfo{booktitle}{\emph{Proceedings of the 2025 CHI Conference on Human Factors in Computing Systems}}. \bibinfo{pages}{1--26}.
\newblock


\bibitem[Cai and Vasconcelos(2018)]%
        {cai2018cascade}
\bibfield{author}{\bibinfo{person}{Zhaowei Cai} {and} \bibinfo{person}{Nuno Vasconcelos}.} \bibinfo{year}{2018}\natexlab{}.
\newblock \showarticletitle{Cascade r-cnn: Delving into high quality object detection}. In \bibinfo{booktitle}{\emph{Proceedings of the IEEE conference on computer vision and pattern recognition}}. \bibinfo{pages}{6154--6162}.
\newblock


\bibitem[Carion et~al\mbox{.}(2020)]%
        {carion2020end}
\bibfield{author}{\bibinfo{person}{Nicolas Carion}, \bibinfo{person}{Francisco Massa}, \bibinfo{person}{Gabriel Synnaeve}, \bibinfo{person}{Nicolas Usunier}, \bibinfo{person}{Alexander Kirillov}, {and} \bibinfo{person}{Sergey Zagoruyko}.} \bibinfo{year}{2020}\natexlab{}.
\newblock \showarticletitle{End-toend object detection with transformers. In eccv}.
\newblock \bibinfo{journal}{\emph{Springer}} \bibinfo{volume}{1}, \bibinfo{number}{2} (\bibinfo{year}{2020}), \bibinfo{pages}{4}.
\newblock


\bibitem[Cesqui et~al\mbox{.}(2013)]%
        {cesqui2013novel}
\bibfield{author}{\bibinfo{person}{Benedetta Cesqui}, \bibinfo{person}{Rolf van De~Langenberg}, \bibinfo{person}{Francesco Lacquaniti}, {and} \bibinfo{person}{Andrea d'Avella}.} \bibinfo{year}{2013}\natexlab{}.
\newblock \showarticletitle{A novel method for measuring gaze orientation in space in unrestrained head conditions}.
\newblock \bibinfo{journal}{\emph{Journal of vision}} \bibinfo{volume}{13}, \bibinfo{number}{8} (\bibinfo{year}{2013}), \bibinfo{pages}{28--28}.
\newblock


\bibitem[Chen(2017)]%
        {chen2017rethinking}
\bibfield{author}{\bibinfo{person}{Liang-Chieh Chen}.} \bibinfo{year}{2017}\natexlab{}.
\newblock \showarticletitle{Rethinking atrous convolution for semantic image segmentation}.
\newblock \bibinfo{journal}{\emph{arXiv preprint arXiv:1706.05587}} (\bibinfo{year}{2017}).
\newblock


\bibitem[Chen et~al\mbox{.}(2017)]%
        {chen2017deeplab}
\bibfield{author}{\bibinfo{person}{Liang-Chieh Chen}, \bibinfo{person}{George Papandreou}, \bibinfo{person}{Iasonas Kokkinos}, \bibinfo{person}{Kevin Murphy}, {and} \bibinfo{person}{Alan~L Yuille}.} \bibinfo{year}{2017}\natexlab{}.
\newblock \showarticletitle{Deeplab: Semantic image segmentation with deep convolutional nets, atrous convolution, and fully connected crfs}.
\newblock \bibinfo{journal}{\emph{IEEE transactions on pattern analysis and machine intelligence}} \bibinfo{volume}{40}, \bibinfo{number}{4} (\bibinfo{year}{2017}), \bibinfo{pages}{834--848}.
\newblock


\bibitem[Chen et~al\mbox{.}(2020)]%
        {Chen2020AugmentingSV}
\bibfield{author}{\bibinfo{person}{Zhutian Chen}, \bibinfo{person}{Wai-Shun Tong}, \bibinfo{person}{Qianwen Wang}, \bibinfo{person}{Benjamin Bach}, {and} \bibinfo{person}{Huamin Qu}.} \bibinfo{year}{2020}\natexlab{}.
\newblock \showarticletitle{Augmenting Static Visualizations with PapARVis Designer}.
\newblock \bibinfo{journal}{\emph{Proceedings of the 2020 CHI Conference on Human Factors in Computing Systems}} (\bibinfo{year}{2020}).
\newblock
\urldef\tempurl%
\url{https://api.semanticscholar.org/CorpusID:218483267}
\showURL{%
\tempurl}


\bibitem[Chidambaram et~al\mbox{.}(2022)]%
        {Chidambaram2022EditARAD}
\bibfield{author}{\bibinfo{person}{Subramanian Chidambaram}, \bibinfo{person}{Sai~Swarup Reddy}, \bibinfo{person}{Matthew Rumple}, \bibinfo{person}{Ananya Ipsita}, \bibinfo{person}{Ana~M. Villanueva}, \bibinfo{person}{Thomas Redick}, \bibinfo{person}{Wolfgang Stuerzlinger}, {and} \bibinfo{person}{Karthik Ramani}.} \bibinfo{year}{2022}\natexlab{}.
\newblock \showarticletitle{EditAR: A Digital Twin Authoring Environment for Creation of AR/VR and Video Instructions from a Single Demonstration}.
\newblock \bibinfo{journal}{\emph{2022 IEEE International Symposium on Mixed and Augmented Reality (ISMAR)}} (\bibinfo{year}{2022}), \bibinfo{pages}{326--335}.
\newblock
\urldef\tempurl%
\url{https://api.semanticscholar.org/CorpusID:254129085}
\showURL{%
\tempurl}


\bibitem[Cho et~al\mbox{.}(2024)]%
        {Cho2024SonoHapticsAA}
\bibfield{author}{\bibinfo{person}{Hyunsung Cho}, \bibinfo{person}{Naveen Sendhilnathan}, \bibinfo{person}{Michael Nebeling}, \bibinfo{person}{Tianyi Wang}, \bibinfo{person}{Purnima Padmanabhan}, \bibinfo{person}{Jonathan Browder}, \bibinfo{person}{David Lindlbauer}, \bibinfo{person}{Tanya~R. Jonker}, {and} \bibinfo{person}{Kashyap Todi}.} \bibinfo{year}{2024}\natexlab{}.
\newblock \showarticletitle{SonoHaptics: An Audio-Haptic Cursor for Gaze-Based Object Selection in XR}.
\newblock
\urldef\tempurl%
\url{https://api.semanticscholar.org/CorpusID:272368428}
\showURL{%
\tempurl}


\bibitem[Dai et~al\mbox{.}(2021)]%
        {dai2021up}
\bibfield{author}{\bibinfo{person}{Zhigang Dai}, \bibinfo{person}{Bolun Cai}, \bibinfo{person}{Yugeng Lin}, {and} \bibinfo{person}{Junying Chen}.} \bibinfo{year}{2021}\natexlab{}.
\newblock \showarticletitle{Up-detr: Unsupervised pre-training for object detection with transformers}. In \bibinfo{booktitle}{\emph{Proceedings of the IEEE/CVF conference on computer vision and pattern recognition}}. \bibinfo{pages}{1601--1610}.
\newblock


\bibitem[Deng et~al\mbox{.}(2017)]%
        {deng2017understanding}
\bibfield{author}{\bibinfo{person}{Shujie Deng}, \bibinfo{person}{Nan Jiang}, \bibinfo{person}{Jian Chang}, \bibinfo{person}{Shihui Guo}, {and} \bibinfo{person}{Jian~J Zhang}.} \bibinfo{year}{2017}\natexlab{}.
\newblock \showarticletitle{Understanding the impact of multimodal interaction using gaze informed mid-air gesture control in 3D virtual objects manipulation}.
\newblock \bibinfo{journal}{\emph{International Journal of Human-Computer Studies}}  \bibinfo{volume}{105} (\bibinfo{year}{2017}), \bibinfo{pages}{68--80}.
\newblock


\bibitem[Dogan(2024)]%
        {Dogan2024UbiquitousMD}
\bibfield{author}{\bibinfo{person}{Mustafa~Doga Dogan}.} \bibinfo{year}{2024}\natexlab{}.
\newblock \showarticletitle{Ubiquitous Metadata: Design and Fabrication of Embedded Markers for Real-World Object Identification and Interaction}.
\newblock \bibinfo{journal}{\emph{ArXiv}}  \bibinfo{volume}{abs/2407.11748} (\bibinfo{year}{2024}).
\newblock
\urldef\tempurl%
\url{https://api.semanticscholar.org/CorpusID:271218769}
\showURL{%
\tempurl}


\bibitem[Dogan et~al\mbox{.}(2024)]%
        {Dogan2024AugmentedOI}
\bibfield{author}{\bibinfo{person}{Mustafa~Doga Dogan}, \bibinfo{person}{Eric~J. Gonzalez}, \bibinfo{person}{Karan Ahuja}, \bibinfo{person}{Ruofei Du}, \bibinfo{person}{Andrea Colacco}, \bibinfo{person}{Johnny Lee}, \bibinfo{person}{Mar Gonz{\'a}lez-Franco}, {and} \bibinfo{person}{David Kim}.} \bibinfo{year}{2024}\natexlab{}.
\newblock \showarticletitle{Augmented Object Intelligence with XR-Objects}.
\newblock
\urldef\tempurl%
\url{https://api.semanticscholar.org/CorpusID:269293799}
\showURL{%
\tempurl}


\bibitem[Drewes and Schmidt(2007)]%
        {drewes2007interacting}
\bibfield{author}{\bibinfo{person}{Heiko Drewes} {and} \bibinfo{person}{Albrecht Schmidt}.} \bibinfo{year}{2007}\natexlab{}.
\newblock \showarticletitle{Interacting with the computer using gaze gestures}. In \bibinfo{booktitle}{\emph{Human-Computer Interaction--INTERACT 2007: 11th IFIP TC 13 International Conference, Rio de Janeiro, Brazil, September 10-14, 2007, Proceedings, Part II 11}}. Springer, \bibinfo{pages}{475--488}.
\newblock


\bibitem[Du et~al\mbox{.}(2022)]%
        {du2022learning}
\bibfield{author}{\bibinfo{person}{Yu Du}, \bibinfo{person}{Fangyun Wei}, \bibinfo{person}{Zihe Zhang}, \bibinfo{person}{Miaojing Shi}, \bibinfo{person}{Yue Gao}, {and} \bibinfo{person}{Guoqi Li}.} \bibinfo{year}{2022}\natexlab{}.
\newblock \showarticletitle{Learning to prompt for open-vocabulary object detection with vision-language model}. In \bibinfo{booktitle}{\emph{Proceedings of the IEEE/CVF Conference on Computer Vision and Pattern Recognition}}. \bibinfo{pages}{14084--14093}.
\newblock


\bibitem[Esteves et~al\mbox{.}(2017)]%
        {esteves2017smoothmoves}
\bibfield{author}{\bibinfo{person}{Augusto Esteves}, \bibinfo{person}{David Verweij}, \bibinfo{person}{Liza Suraiya}, \bibinfo{person}{Rasel Islam}, \bibinfo{person}{Youryang Lee}, {and} \bibinfo{person}{Ian Oakley}.} \bibinfo{year}{2017}\natexlab{}.
\newblock \showarticletitle{Smoothmoves: Smooth pursuits head movements for augmented reality}. In \bibinfo{booktitle}{\emph{Proceedings of the 30th annual acm symposium on user interface software and technology}}. \bibinfo{pages}{167--178}.
\newblock


\bibitem[Fashimpaur et~al\mbox{.}(2025)]%
        {squiggle2025}
\bibfield{author}{\bibinfo{person}{Jacqui Fashimpaur}, \bibinfo{person}{Tovi Grossman}, \bibinfo{person}{Ben Lafreniere}, \bibinfo{person}{Naveen Sendhilnathan}, \bibinfo{person}{Kashyap Todi}, \bibinfo{person}{Tianyi Wang}, \bibinfo{person}{Ting Zhang}, {and} \bibinfo{person}{Tanya~R. Jonker}.} \bibinfo{year}{2025}\natexlab{}.
\newblock \showarticletitle{Squiggle: Multimodal Lasso Selection in the Real World}. In \bibinfo{booktitle}{\emph{Proceedings of the 38th Annual ACM Symposium on User Interface Software and Technology}} \emph{(\bibinfo{series}{UIST '25})}. \bibinfo{publisher}{Association for Computing Machinery}, \bibinfo{address}{New York, NY, USA}, Article \bibinfo{articleno}{115}, \bibinfo{numpages}{16}~pages.
\newblock
\showISBNx{9798400720376}
\urldef\tempurl%
\url{https://doi.org/10.1145/3746059.3747684}
\showDOI{\tempurl}


\bibitem[Feng et~al\mbox{.}(2022)]%
        {feng2022promptdet}
\bibfield{author}{\bibinfo{person}{Chengjian Feng}, \bibinfo{person}{Yujie Zhong}, \bibinfo{person}{Zequn Jie}, \bibinfo{person}{Xiangxiang Chu}, \bibinfo{person}{Haibing Ren}, \bibinfo{person}{Xiaolin Wei}, \bibinfo{person}{Weidi Xie}, {and} \bibinfo{person}{Lin Ma}.} \bibinfo{year}{2022}\natexlab{}.
\newblock \showarticletitle{Promptdet: Towards open-vocabulary detection using uncurated images}. In \bibinfo{booktitle}{\emph{European Conference on Computer Vision}}. Springer, \bibinfo{pages}{701--717}.
\newblock


\bibitem[Fernandes et~al\mbox{.}(2023)]%
        {fernandes2023leveling}
\bibfield{author}{\bibinfo{person}{Ajoy~S Fernandes}, \bibinfo{person}{T~Scott Murdison}, {and} \bibinfo{person}{Michael~J Proulx}.} \bibinfo{year}{2023}\natexlab{}.
\newblock \showarticletitle{Leveling the playing field: A comparative reevaluation of unmodified eye tracking as an input and interaction modality for VR}.
\newblock \bibinfo{journal}{\emph{IEEE Transactions on Visualization and Computer Graphics}} \bibinfo{volume}{29}, \bibinfo{number}{5} (\bibinfo{year}{2023}), \bibinfo{pages}{2269--2279}.
\newblock


\bibitem[Fernandes et~al\mbox{.}(2024)]%
        {fernandes2024degraded}
\bibfield{author}{\bibinfo{person}{Ajoy~Savio Fernandes}, \bibinfo{person}{T.~Scott Murdison}, \bibinfo{person}{Immo Schuetz}, \bibinfo{person}{Oleg Komogortsev}, {and} \bibinfo{person}{Michael~J Proulx}.} \bibinfo{year}{2024}\natexlab{}.
\newblock \showarticletitle{The Effect of Degraded Eye Tracking Accuracy on Interactions in VR}. In \bibinfo{booktitle}{\emph{Proceedings of the 2024 Symposium on Eye Tracking Research and Applications}} (Glasgow, United Kingdom) \emph{(\bibinfo{series}{ETRA '24})}. \bibinfo{publisher}{Association for Computing Machinery}, \bibinfo{address}{New York, NY, USA}, Article \bibinfo{articleno}{63}, \bibinfo{numpages}{7}~pages.
\newblock
\showISBNx{9798400706073}
\urldef\tempurl%
\url{https://doi.org/10.1145/3649902.3656369}
\showDOI{\tempurl}


\bibitem[Giannopoulos et~al\mbox{.}(2015)]%
        {giannopoulos2015gazenav}
\bibfield{author}{\bibinfo{person}{Ioannis Giannopoulos}, \bibinfo{person}{Peter Kiefer}, {and} \bibinfo{person}{Martin Raubal}.} \bibinfo{year}{2015}\natexlab{}.
\newblock \showarticletitle{GazeNav: Gaze-based pedestrian navigation}. In \bibinfo{booktitle}{\emph{Proceedings of the 17th international conference on human-computer interaction with mobile devices and services}}. \bibinfo{pages}{337--346}.
\newblock


\bibitem[Girshick et~al\mbox{.}(2014)]%
        {girshick2014rich}
\bibfield{author}{\bibinfo{person}{Ross Girshick}, \bibinfo{person}{Jeff Donahue}, \bibinfo{person}{Trevor Darrell}, {and} \bibinfo{person}{Jitendra Malik}.} \bibinfo{year}{2014}\natexlab{}.
\newblock \showarticletitle{Rich feature hierarchies for accurate object detection and semantic segmentation}. In \bibinfo{booktitle}{\emph{Proceedings of the IEEE conference on computer vision and pattern recognition}}. \bibinfo{pages}{580--587}.
\newblock


\bibitem[Gu et~al\mbox{.}(2021)]%
        {gu2021open}
\bibfield{author}{\bibinfo{person}{Xiuye Gu}, \bibinfo{person}{Tsung-Yi Lin}, \bibinfo{person}{Weicheng Kuo}, {and} \bibinfo{person}{Yin Cui}.} \bibinfo{year}{2021}\natexlab{}.
\newblock \showarticletitle{Open-vocabulary object detection via vision and language knowledge distillation}.
\newblock \bibinfo{journal}{\emph{arXiv preprint arXiv:2104.13921}} (\bibinfo{year}{2021}).
\newblock


\bibitem[Hansen et~al\mbox{.}(2018)]%
        {hansen2018fitts}
\bibfield{author}{\bibinfo{person}{John~Paulin Hansen}, \bibinfo{person}{Vijay Rajanna}, \bibinfo{person}{I~Scott MacKenzie}, {and} \bibinfo{person}{Per B{\ae}kgaard}.} \bibinfo{year}{2018}\natexlab{}.
\newblock \showarticletitle{A Fitts' law study of click and dwell interaction by gaze, head and mouse with a head-mounted display}. In \bibinfo{booktitle}{\emph{Proceedings of the Workshop on Communication by Gaze Interaction}}. \bibinfo{pages}{1--5}.
\newblock


\bibitem[Hart(2006)]%
        {Hart2006NasaTaskLI}
\bibfield{author}{\bibinfo{person}{S.~G. Hart}.} \bibinfo{year}{2006}\natexlab{}.
\newblock \showarticletitle{Nasa-Task Load Index (NASA-TLX); 20 Years Later}.
\newblock \bibinfo{journal}{\emph{Proceedings of the Human Factors and Ergonomics Society Annual Meeting}}  \bibinfo{volume}{50} (\bibinfo{year}{2006}), \bibinfo{pages}{904 -- 908}.
\newblock
\urldef\tempurl%
\url{https://api.semanticscholar.org/CorpusID:6292200}
\showURL{%
\tempurl}


\bibitem[He et~al\mbox{.}(2017)]%
        {he2017mask}
\bibfield{author}{\bibinfo{person}{Kaiming He}, \bibinfo{person}{Georgia Gkioxari}, \bibinfo{person}{Piotr Doll{\'a}r}, {and} \bibinfo{person}{Ross Girshick}.} \bibinfo{year}{2017}\natexlab{}.
\newblock \showarticletitle{Mask r-cnn}. In \bibinfo{booktitle}{\emph{Proceedings of the IEEE international conference on computer vision}}. \bibinfo{pages}{2961--2969}.
\newblock


\bibitem[Henderson and Feiner(2008)]%
        {Henderson2008OpportunisticCL}
\bibfield{author}{\bibinfo{person}{Steven~J. Henderson} {and} \bibinfo{person}{Steven~K. Feiner}.} \bibinfo{year}{2008}\natexlab{}.
\newblock \showarticletitle{Opportunistic controls: leveraging natural affordances as tangible user interfaces for augmented reality}. In \bibinfo{booktitle}{\emph{Virtual Reality Software and Technology}}.
\newblock
\urldef\tempurl%
\url{https://api.semanticscholar.org/CorpusID:11199619}
\showURL{%
\tempurl}


\bibitem[Heun et~al\mbox{.}(2013)]%
        {Heun2013RealityEP}
\bibfield{author}{\bibinfo{person}{Valentin Heun}, \bibinfo{person}{James Hobin}, {and} \bibinfo{person}{Pattie Maes}.} \bibinfo{year}{2013}\natexlab{}.
\newblock \showarticletitle{Reality editor: programming smarter objects}.
\newblock \bibinfo{journal}{\emph{Proceedings of the 2013 ACM conference on Pervasive and ubiquitous computing adjunct publication}} (\bibinfo{year}{2013}).
\newblock
\urldef\tempurl%
\url{https://api.semanticscholar.org/CorpusID:14748376}
\showURL{%
\tempurl}


\bibitem[Hyrskykari et~al\mbox{.}(2012)]%
        {hyrskykari2012gaze}
\bibfield{author}{\bibinfo{person}{Aulikki Hyrskykari}, \bibinfo{person}{Howell Istance}, {and} \bibinfo{person}{Stephen Vickers}.} \bibinfo{year}{2012}\natexlab{}.
\newblock \showarticletitle{Gaze gestures or dwell-based interaction?}. In \bibinfo{booktitle}{\emph{Proceedings of the Symposium on Eye Tracking Research and Applications}}. \bibinfo{pages}{229--232}.
\newblock


\bibitem[Istance et~al\mbox{.}(2010)]%
        {istance2010designing}
\bibfield{author}{\bibinfo{person}{Howell Istance}, \bibinfo{person}{Aulikki Hyrskykari}, \bibinfo{person}{Lauri Immonen}, \bibinfo{person}{Santtu Mansikkamaa}, {and} \bibinfo{person}{Stephen Vickers}.} \bibinfo{year}{2010}\natexlab{}.
\newblock \showarticletitle{Designing gaze gestures for gaming: an investigation of performance}. In \bibinfo{booktitle}{\emph{Proceedings of the 2010 Symposium on Eye-Tracking Research \& Applications}}. \bibinfo{pages}{323--330}.
\newblock


\bibitem[Jacob(1990)]%
        {jacob1990you}
\bibfield{author}{\bibinfo{person}{Robert~JK Jacob}.} \bibinfo{year}{1990}\natexlab{}.
\newblock \showarticletitle{What you look at is what you get: eye movement-based interaction techniques}. In \bibinfo{booktitle}{\emph{Proceedings of the SIGCHI conference on Human factors in computing systems}}. \bibinfo{pages}{11--18}.
\newblock


\bibitem[Jang et~al\mbox{.}(2017)]%
        {Jang2017MetaphoricHG}
\bibfield{author}{\bibinfo{person}{Youngkyoon Jang}, \bibinfo{person}{Ikbeom Jeon}, \bibinfo{person}{Tae-Kyun Kim}, {and} \bibinfo{person}{Woontack Woo}.} \bibinfo{year}{2017}\natexlab{}.
\newblock \showarticletitle{Metaphoric Hand Gestures for Orientation-Aware VR Object Manipulation With an Egocentric Viewpoint}.
\newblock \bibinfo{journal}{\emph{IEEE Transactions on Human-Machine Systems}}  \bibinfo{volume}{47} (\bibinfo{year}{2017}), \bibinfo{pages}{113--127}.
\newblock
\urldef\tempurl%
\url{https://api.semanticscholar.org/CorpusID:24522968}
\showURL{%
\tempurl}


\bibitem[Joseph et~al\mbox{.}(2021)]%
        {joseph2021towards}
\bibfield{author}{\bibinfo{person}{KJ Joseph}, \bibinfo{person}{Salman Khan}, \bibinfo{person}{Fahad~Shahbaz Khan}, {and} \bibinfo{person}{Vineeth~N Balasubramanian}.} \bibinfo{year}{2021}\natexlab{}.
\newblock \showarticletitle{Towards open world object detection}. In \bibinfo{booktitle}{\emph{Proceedings of the IEEE/CVF conference on computer vision and pattern recognition}}. \bibinfo{pages}{5830--5840}.
\newblock


\bibitem[Khamis et~al\mbox{.}(2017)]%
        {khamis2017eyescout}
\bibfield{author}{\bibinfo{person}{Mohamed Khamis}, \bibinfo{person}{Axel Hoesl}, \bibinfo{person}{Alexander Klimczak}, \bibinfo{person}{Martin Reiss}, \bibinfo{person}{Florian Alt}, {and} \bibinfo{person}{Andreas Bulling}.} \bibinfo{year}{2017}\natexlab{}.
\newblock \showarticletitle{Eyescout: Active eye tracking for position and movement independent gaze interaction with large public displays}. In \bibinfo{booktitle}{\emph{Proceedings of the 30th annual ACM symposium on user interface software and technology}}. \bibinfo{pages}{155--166}.
\newblock


\bibitem[Kirillov et~al\mbox{.}(2019)]%
        {kirillov2019panoptic}
\bibfield{author}{\bibinfo{person}{Alexander Kirillov}, \bibinfo{person}{Kaiming He}, \bibinfo{person}{Ross Girshick}, \bibinfo{person}{Carsten Rother}, {and} \bibinfo{person}{Piotr Doll{\'a}r}.} \bibinfo{year}{2019}\natexlab{}.
\newblock \showarticletitle{Panoptic segmentation}. In \bibinfo{booktitle}{\emph{Proceedings of the IEEE/CVF conference on computer vision and pattern recognition}}. \bibinfo{pages}{9404--9413}.
\newblock


\bibitem[Kirillov et~al\mbox{.}(2023)]%
        {kirillov2023segment}
\bibfield{author}{\bibinfo{person}{Alexander Kirillov}, \bibinfo{person}{Eric Mintun}, \bibinfo{person}{Nikhila Ravi}, \bibinfo{person}{Hanzi Mao}, \bibinfo{person}{Chloe Rolland}, \bibinfo{person}{Laura Gustafson}, \bibinfo{person}{Tete Xiao}, \bibinfo{person}{Spencer Whitehead}, \bibinfo{person}{Alexander~C Berg}, \bibinfo{person}{Wan-Yen Lo}, {et~al\mbox{.}}} \bibinfo{year}{2023}\natexlab{}.
\newblock \showarticletitle{Segment anything}. In \bibinfo{booktitle}{\emph{Proceedings of the IEEE/CVF International Conference on Computer Vision}}. \bibinfo{pages}{4015--4026}.
\newblock


\bibitem[Konrad et~al\mbox{.}(2024)]%
        {konrad2024gazegpt}
\bibfield{author}{\bibinfo{person}{Robert Konrad}, \bibinfo{person}{Nitish Padmanaban}, \bibinfo{person}{J~Gabriel Buckmaster}, \bibinfo{person}{Kevin~C Boyle}, {and} \bibinfo{person}{Gordon Wetzstein}.} \bibinfo{year}{2024}\natexlab{}.
\newblock \showarticletitle{Gazegpt: Augmenting human capabilities using gaze-contingent contextual ai for smart eyewear}.
\newblock \bibinfo{journal}{\emph{arXiv preprint arXiv:2401.17217}} (\bibinfo{year}{2024}).
\newblock


\bibitem[Kwok et~al\mbox{.}(2019)]%
        {kwok2019gaze}
\bibfield{author}{\bibinfo{person}{Tiffany~CK Kwok}, \bibinfo{person}{Peter Kiefer}, \bibinfo{person}{Victor~R Schinazi}, \bibinfo{person}{Benjamin Adams}, {and} \bibinfo{person}{Martin Raubal}.} \bibinfo{year}{2019}\natexlab{}.
\newblock \showarticletitle{Gaze-guided narratives: Adapting audio guide content to gaze in virtual and real environments}. In \bibinfo{booktitle}{\emph{Proceedings of the 2019 CHI Conference on Human Factors in Computing Systems}}. \bibinfo{pages}{1--12}.
\newblock


\bibitem[Kyt{\"o} et~al\mbox{.}(2018)]%
        {kyto2018pinpointing}
\bibfield{author}{\bibinfo{person}{Mikko Kyt{\"o}}, \bibinfo{person}{Barrett Ens}, \bibinfo{person}{Thammathip Piumsomboon}, \bibinfo{person}{Gun~A Lee}, {and} \bibinfo{person}{Mark Billinghurst}.} \bibinfo{year}{2018}\natexlab{}.
\newblock \showarticletitle{Pinpointing: Precise head-and eye-based target selection for augmented reality}. In \bibinfo{booktitle}{\emph{Proceedings of the 2018 CHI Conference on Human Factors in Computing Systems}}. \bibinfo{pages}{1--14}.
\newblock


\bibitem[Larsson et~al\mbox{.}(2016)]%
        {larsson2016head}
\bibfield{author}{\bibinfo{person}{Linn{\'e}a Larsson}, \bibinfo{person}{Andrea Schwaller}, \bibinfo{person}{Marcus Nystr{\"o}m}, {and} \bibinfo{person}{Martin Stridh}.} \bibinfo{year}{2016}\natexlab{}.
\newblock \showarticletitle{Head movement compensation and multi-modal event detection in eye-tracking data for unconstrained head movements}.
\newblock \bibinfo{journal}{\emph{Journal of neuroscience methods}}  \bibinfo{volume}{274} (\bibinfo{year}{2016}), \bibinfo{pages}{13--26}.
\newblock


\bibitem[Lee et~al\mbox{.}(2021)]%
        {Lee2021WhatsTA}
\bibfield{author}{\bibinfo{person}{Jaewook Lee}, \bibinfo{person}{Sebasti{\'a}n~Sanhueza Rodr{\'i}guez}, \bibinfo{person}{Raahul Natarrajan}, \bibinfo{person}{Jacqueline Chen}, \bibinfo{person}{Harsh Deep}, {and} \bibinfo{person}{Alex Kirlik}.} \bibinfo{year}{2021}\natexlab{}.
\newblock \showarticletitle{What’s This? A Voice and Touch Multimodal Approach for Ambiguity Resolution in Voice Assistants}.
\newblock \bibinfo{journal}{\emph{Proceedings of the 2021 International Conference on Multimodal Interaction}} (\bibinfo{year}{2021}).
\newblock
\urldef\tempurl%
\url{https://api.semanticscholar.org/CorpusID:238992795}
\showURL{%
\tempurl}


\bibitem[Lee et~al\mbox{.}(2024)]%
        {lee2024gazepointar}
\bibfield{author}{\bibinfo{person}{Jaewook Lee}, \bibinfo{person}{Jun Wang}, \bibinfo{person}{Elizabeth Brown}, \bibinfo{person}{Liam Chu}, \bibinfo{person}{Sebastian~S Rodriguez}, {and} \bibinfo{person}{Jon~E Froehlich}.} \bibinfo{year}{2024}\natexlab{}.
\newblock \showarticletitle{GazePointAR: A Context-Aware Multimodal Voice Assistant for Pronoun Disambiguation in Wearable Augmented Reality}. In \bibinfo{booktitle}{\emph{Proceedings of the CHI Conference on Human Factors in Computing Systems}}. \bibinfo{pages}{1--20}.
\newblock


\bibitem[Li et~al\mbox{.}(2023)]%
        {li2023mask}
\bibfield{author}{\bibinfo{person}{Feng Li}, \bibinfo{person}{Hao Zhang}, \bibinfo{person}{Huaizhe Xu}, \bibinfo{person}{Shilong Liu}, \bibinfo{person}{Lei Zhang}, \bibinfo{person}{Lionel~M Ni}, {and} \bibinfo{person}{Heung-Yeung Shum}.} \bibinfo{year}{2023}\natexlab{}.
\newblock \showarticletitle{Mask dino: Towards a unified transformer-based framework for object detection and segmentation}. In \bibinfo{booktitle}{\emph{Proceedings of the IEEE/CVF Conference on Computer Vision and Pattern Recognition}}. \bibinfo{pages}{3041--3050}.
\newblock


\bibitem[Lin et~al\mbox{.}(2017)]%
        {lin2017focal}
\bibfield{author}{\bibinfo{person}{Tsung-Yi Lin}, \bibinfo{person}{Priya Goyal}, \bibinfo{person}{Ross Girshick}, \bibinfo{person}{Kaiming He}, {and} \bibinfo{person}{Piotr Doll{\'a}r}.} \bibinfo{year}{2017}\natexlab{}.
\newblock \showarticletitle{Focal loss for dense object detection}. In \bibinfo{booktitle}{\emph{Proceedings of the IEEE international conference on computer vision}}. \bibinfo{pages}{2980--2988}.
\newblock


\bibitem[Liu et~al\mbox{.}(2022)]%
        {Liu2022SimpleClickII}
\bibfield{author}{\bibinfo{person}{Qin Liu}, \bibinfo{person}{Zhenlin Xu}, \bibinfo{person}{Gedas Bertasius}, {and} \bibinfo{person}{Marc Niethammer}.} \bibinfo{year}{2022}\natexlab{}.
\newblock \showarticletitle{SimpleClick: Interactive Image Segmentation with Simple Vision Transformers}.
\newblock \bibinfo{journal}{\emph{2023 IEEE/CVF International Conference on Computer Vision (ICCV)}} (\bibinfo{year}{2022}), \bibinfo{pages}{22233--22243}.
\newblock
\urldef\tempurl%
\url{https://api.semanticscholar.org/CorpusID:253018432}
\showURL{%
\tempurl}


\bibitem[Liu et~al\mbox{.}(2023)]%
        {Liu2023GroundingDM}
\bibfield{author}{\bibinfo{person}{Shilong Liu}, \bibinfo{person}{Zhaoyang Zeng}, \bibinfo{person}{Tianhe Ren}, \bibinfo{person}{Feng Li}, \bibinfo{person}{Hao Zhang}, \bibinfo{person}{Jie Yang}, \bibinfo{person}{Chun yue Li}, \bibinfo{person}{Jianwei Yang}, \bibinfo{person}{Hang Su}, \bibinfo{person}{Jun-Juan Zhu}, {and} \bibinfo{person}{Lei Zhang}.} \bibinfo{year}{2023}\natexlab{}.
\newblock \showarticletitle{Grounding DINO: Marrying DINO with Grounded Pre-Training for Open-Set Object Detection}.
\newblock \bibinfo{journal}{\emph{ArXiv}}  \bibinfo{volume}{abs/2303.05499} (\bibinfo{year}{2023}).
\newblock
\urldef\tempurl%
\url{https://api.semanticscholar.org/CorpusID:257427307}
\showURL{%
\tempurl}


\bibitem[Liu et~al\mbox{.}(2016)]%
        {liu2016ssd}
\bibfield{author}{\bibinfo{person}{Wei Liu}, \bibinfo{person}{Dragomir Anguelov}, \bibinfo{person}{Dumitru Erhan}, \bibinfo{person}{Christian Szegedy}, \bibinfo{person}{Scott Reed}, \bibinfo{person}{Cheng-Yang Fu}, {and} \bibinfo{person}{Alexander~C Berg}.} \bibinfo{year}{2016}\natexlab{}.
\newblock \showarticletitle{Ssd: Single shot multibox detector}. In \bibinfo{booktitle}{\emph{Computer Vision--ECCV 2016: 14th European Conference, Amsterdam, The Netherlands, October 11--14, 2016, Proceedings, Part I 14}}. Springer, \bibinfo{pages}{21--37}.
\newblock


\bibitem[L{\"o}cken et~al\mbox{.}(2012)]%
        {locken2012user}
\bibfield{author}{\bibinfo{person}{Andreas L{\"o}cken}, \bibinfo{person}{Tobias Hesselmann}, \bibinfo{person}{Martin Pielot}, \bibinfo{person}{Niels Henze}, {and} \bibinfo{person}{Susanne Boll}.} \bibinfo{year}{2012}\natexlab{}.
\newblock \showarticletitle{User-centred process for the definition of free-hand gestures applied to controlling music playback}.
\newblock \bibinfo{journal}{\emph{Multimedia systems}}  \bibinfo{volume}{18} (\bibinfo{year}{2012}), \bibinfo{pages}{15--31}.
\newblock


\bibitem[Long et~al\mbox{.}(2015)]%
        {long2015fully}
\bibfield{author}{\bibinfo{person}{Jonathan Long}, \bibinfo{person}{Evan Shelhamer}, {and} \bibinfo{person}{Trevor Darrell}.} \bibinfo{year}{2015}\natexlab{}.
\newblock \showarticletitle{Fully convolutional networks for semantic segmentation}. In \bibinfo{booktitle}{\emph{Proceedings of the IEEE conference on computer vision and pattern recognition}}. \bibinfo{pages}{3431--3440}.
\newblock


\bibitem[Lu et~al\mbox{.}(2021)]%
        {lu2021exploration}
\bibfield{author}{\bibinfo{person}{Feiyu Lu}, \bibinfo{person}{Shakiba Davari}, {and} \bibinfo{person}{Doug Bowman}.} \bibinfo{year}{2021}\natexlab{}.
\newblock \showarticletitle{Exploration of techniques for rapid activation of glanceable information in head-worn augmented reality}. In \bibinfo{booktitle}{\emph{Proceedings of the 2021 ACM Symposium on Spatial User Interaction}}. \bibinfo{pages}{1--11}.
\newblock


\bibitem[Lv et~al\mbox{.}(2023)]%
        {Lv2023DETRsBY}
\bibfield{author}{\bibinfo{person}{Wenyu Lv}, \bibinfo{person}{Shangliang Xu}, \bibinfo{person}{Yian Zhao}, \bibinfo{person}{Guanzhong Wang}, \bibinfo{person}{Jinman Wei}, \bibinfo{person}{Cheng Cui}, \bibinfo{person}{Yuning Du}, \bibinfo{person}{Qingqing Dang}, {and} \bibinfo{person}{Yi Liu}.} \bibinfo{year}{2023}\natexlab{}.
\newblock \showarticletitle{DETRs Beat YOLOs on Real-time Object Detection}.
\newblock \bibinfo{journal}{\emph{ArXiv}}  \bibinfo{volume}{abs/2304.08069} (\bibinfo{year}{2023}).
\newblock
\urldef\tempurl%
\url{https://api.semanticscholar.org/CorpusID:258179840}
\showURL{%
\tempurl}


\bibitem[Lystb{\ae}k et~al\mbox{.}(2022)]%
        {lystbaek2022gaze}
\bibfield{author}{\bibinfo{person}{Mathias~N Lystb{\ae}k}, \bibinfo{person}{Peter Rosenberg}, \bibinfo{person}{Ken Pfeuffer}, \bibinfo{person}{Jens~Emil Gr{\o}nb{\ae}k}, {and} \bibinfo{person}{Hans Gellersen}.} \bibinfo{year}{2022}\natexlab{}.
\newblock \showarticletitle{Gaze-hand alignment: Combining eye gaze and mid-air pointing for interacting with menus in augmented reality}.
\newblock \bibinfo{journal}{\emph{Proceedings of the ACM on Human-Computer Interaction}} \bibinfo{volume}{6}, \bibinfo{number}{ETRA} (\bibinfo{year}{2022}), \bibinfo{pages}{1--18}.
\newblock


\bibitem[Majaranta and Bulling(2014)]%
        {majaranta2014eye}
\bibfield{author}{\bibinfo{person}{P{\"a}ivi Majaranta} {and} \bibinfo{person}{Andreas Bulling}.} \bibinfo{year}{2014}\natexlab{}.
\newblock \showarticletitle{Eye tracking and eye-based human--computer interaction}.
\newblock In \bibinfo{booktitle}{\emph{Advances in physiological computing}}. \bibinfo{publisher}{Springer}, \bibinfo{pages}{39--65}.
\newblock


\bibitem[Mayer et~al\mbox{.}(2020)]%
        {Mayer2020EnhancingMV}
\bibfield{author}{\bibinfo{person}{Sven Mayer}, \bibinfo{person}{Gierad Laput}, {and} \bibinfo{person}{Chris Harrison}.} \bibinfo{year}{2020}\natexlab{}.
\newblock \showarticletitle{Enhancing Mobile Voice Assistants with WorldGaze}.
\newblock \bibinfo{journal}{\emph{Proceedings of the 2020 CHI Conference on Human Factors in Computing Systems}} (\bibinfo{year}{2020}).
\newblock
\urldef\tempurl%
\url{https://api.semanticscholar.org/CorpusID:215776069}
\showURL{%
\tempurl}


\bibitem[Minderer et~al\mbox{.}(2023)]%
        {Minderer2023ScalingOO}
\bibfield{author}{\bibinfo{person}{Matthias Minderer}, \bibinfo{person}{Alexey~A. Gritsenko}, {and} \bibinfo{person}{Neil Houlsby}.} \bibinfo{year}{2023}\natexlab{}.
\newblock \showarticletitle{Scaling Open-Vocabulary Object Detection}.
\newblock \bibinfo{journal}{\emph{ArXiv}}  \bibinfo{volume}{abs/2306.09683} (\bibinfo{year}{2023}).
\newblock
\urldef\tempurl%
\url{https://api.semanticscholar.org/CorpusID:259187664}
\showURL{%
\tempurl}


\bibitem[Miniotas et~al\mbox{.}(2006)]%
        {Miniotas2006SpeechaugmentedEG}
\bibfield{author}{\bibinfo{person}{Darius Miniotas}, \bibinfo{person}{Oleg Spakov}, \bibinfo{person}{Ivan Tugoy}, {and} \bibinfo{person}{I.~Scott MacKenzie}.} \bibinfo{year}{2006}\natexlab{}.
\newblock \showarticletitle{Speech-augmented eye gaze interaction with small closely spaced targets}.
\newblock \bibinfo{journal}{\emph{Proceedings of the 2006 symposium on Eye tracking research \& applications}} (\bibinfo{year}{2006}).
\newblock
\urldef\tempurl%
\url{https://api.semanticscholar.org/CorpusID:10705855}
\showURL{%
\tempurl}


\bibitem[Monteiro et~al\mbox{.}(2023)]%
        {Monteiro2023TeachableRP}
\bibfield{author}{\bibinfo{person}{Kyzyl Monteiro}, \bibinfo{person}{Ritik Vatsal}, \bibinfo{person}{Neil Chulpongsatorn}, \bibinfo{person}{Aman Parnami}, {and} \bibinfo{person}{Ryo Suzuki}.} \bibinfo{year}{2023}\natexlab{}.
\newblock \showarticletitle{Teachable Reality: Prototyping Tangible Augmented Reality with Everyday Objects by Leveraging Interactive Machine Teaching}.
\newblock \bibinfo{journal}{\emph{Proceedings of the 2023 CHI Conference on Human Factors in Computing Systems}} (\bibinfo{year}{2023}).
\newblock
\urldef\tempurl%
\url{https://api.semanticscholar.org/CorpusID:257078906}
\showURL{%
\tempurl}


\bibitem[Neubeck and Gool(2006)]%
        {Neubeck2006EfficientNS}
\bibfield{author}{\bibinfo{person}{Alexander Neubeck} {and} \bibinfo{person}{Luc~Van Gool}.} \bibinfo{year}{2006}\natexlab{}.
\newblock \showarticletitle{Efficient Non-Maximum Suppression}.
\newblock \bibinfo{journal}{\emph{18th International Conference on Pattern Recognition (ICPR'06)}}  \bibinfo{volume}{3} (\bibinfo{year}{2006}), \bibinfo{pages}{850--855}.
\newblock
\urldef\tempurl%
\url{https://api.semanticscholar.org/CorpusID:5057778}
\showURL{%
\tempurl}


\bibitem[Nielsen(1993)]%
        {nielsen1993noncommand}
\bibfield{author}{\bibinfo{person}{Jakob Nielsen}.} \bibinfo{year}{1993}\natexlab{}.
\newblock \showarticletitle{Noncommand user interfaces}.
\newblock \bibinfo{journal}{\emph{Commun. ACM}} \bibinfo{volume}{36}, \bibinfo{number}{4} (\bibinfo{year}{1993}), \bibinfo{pages}{83--99}.
\newblock


\bibitem[Pai et~al\mbox{.}(2016)]%
        {pai2016transparent}
\bibfield{author}{\bibinfo{person}{Yun~Suen Pai}, \bibinfo{person}{Benjamin Outram}, \bibinfo{person}{Noriyasu Vontin}, {and} \bibinfo{person}{Kai Kunze}.} \bibinfo{year}{2016}\natexlab{}.
\newblock \showarticletitle{Transparent reality: Using eye gaze focus depth as interaction modality}. In \bibinfo{booktitle}{\emph{Adjunct Proceedings of the 29th Annual ACM Symposium on User Interface Software and Technology}}. \bibinfo{pages}{171--172}.
\newblock


\bibitem[Park et~al\mbox{.}(2008)]%
        {park2008wearable}
\bibfield{author}{\bibinfo{person}{Hyung~Min Park}, \bibinfo{person}{Seok~Han Lee}, {and} \bibinfo{person}{Jong~Soo Choi}.} \bibinfo{year}{2008}\natexlab{}.
\newblock \showarticletitle{Wearable augmented reality system using gaze interaction}. In \bibinfo{booktitle}{\emph{2008 7th IEEE/ACM International Symposium on Mixed and Augmented Reality}}. IEEE, \bibinfo{pages}{175--176}.
\newblock


\bibitem[Pfeuffer et~al\mbox{.}(2017)]%
        {pfeuffer2017gaze+}
\bibfield{author}{\bibinfo{person}{Ken Pfeuffer}, \bibinfo{person}{Benedikt Mayer}, \bibinfo{person}{Diako Mardanbegi}, {and} \bibinfo{person}{Hans Gellersen}.} \bibinfo{year}{2017}\natexlab{}.
\newblock \showarticletitle{Gaze+ pinch interaction in virtual reality}. In \bibinfo{booktitle}{\emph{Proceedings of the 5th symposium on spatial user interaction}}. \bibinfo{pages}{99--108}.
\newblock


\bibitem[Piumsomboon et~al\mbox{.}(2017)]%
        {piumsomboon2017exploring}
\bibfield{author}{\bibinfo{person}{Thammathip Piumsomboon}, \bibinfo{person}{Gun Lee}, \bibinfo{person}{Robert~W Lindeman}, {and} \bibinfo{person}{Mark Billinghurst}.} \bibinfo{year}{2017}\natexlab{}.
\newblock \showarticletitle{Exploring natural eye-gaze-based interaction for immersive virtual reality}. In \bibinfo{booktitle}{\emph{2017 IEEE symposium on 3D user interfaces (3DUI)}}. IEEE, \bibinfo{pages}{36--39}.
\newblock


\bibitem[Pu et~al\mbox{.}(2025)]%
        {pu2025promemassist}
\bibfield{author}{\bibinfo{person}{Kevin Pu}, \bibinfo{person}{Ting Zhang}, \bibinfo{person}{Naveen Sendhilnathan}, \bibinfo{person}{Sebastian Freitag}, \bibinfo{person}{Raj Sodhi}, {and} \bibinfo{person}{Tanya~R Jonker}.} \bibinfo{year}{2025}\natexlab{}.
\newblock \showarticletitle{ProMemAssist: Exploring Timely Proactive Assistance Through Working Memory Modeling in Multi-Modal Wearable Devices}. In \bibinfo{booktitle}{\emph{Proceedings of the 38th Annual ACM Symposium on User Interface Software and Technology}}. \bibinfo{pages}{1--19}.
\newblock


\bibitem[Qian and Teather(2017)]%
        {qian2017eyes}
\bibfield{author}{\bibinfo{person}{Yuan~Yuan Qian} {and} \bibinfo{person}{Robert~J Teather}.} \bibinfo{year}{2017}\natexlab{}.
\newblock \showarticletitle{The eyes don't have it: an empirical comparison of head-based and eye-based selection in virtual reality}. In \bibinfo{booktitle}{\emph{Proceedings of the 5th Symposium on Spatial User Interaction}}. \bibinfo{pages}{91--98}.
\newblock


\bibitem[Rajanna and Hansen(2018)]%
        {rajanna2018gaze}
\bibfield{author}{\bibinfo{person}{Vijay Rajanna} {and} \bibinfo{person}{John~Paulin Hansen}.} \bibinfo{year}{2018}\natexlab{}.
\newblock \showarticletitle{Gaze typing in virtual reality: impact of keyboard design, selection method, and motion}. In \bibinfo{booktitle}{\emph{Proceedings of the 2018 ACM symposium on eye tracking research \& applications}}. \bibinfo{pages}{1--10}.
\newblock


\bibitem[Rasheed et~al\mbox{.}(2024)]%
        {rasheed2024glamm}
\bibfield{author}{\bibinfo{person}{Hanoona Rasheed}, \bibinfo{person}{Muhammad Maaz}, \bibinfo{person}{Sahal Shaji}, \bibinfo{person}{Abdelrahman Shaker}, \bibinfo{person}{Salman Khan}, \bibinfo{person}{Hisham Cholakkal}, \bibinfo{person}{Rao~M Anwer}, \bibinfo{person}{Eric Xing}, \bibinfo{person}{Ming-Hsuan Yang}, {and} \bibinfo{person}{Fahad~S Khan}.} \bibinfo{year}{2024}\natexlab{}.
\newblock \showarticletitle{Glamm: Pixel grounding large multimodal model}. In \bibinfo{booktitle}{\emph{Proceedings of the IEEE/CVF Conference on Computer Vision and Pattern Recognition}}. \bibinfo{pages}{13009--13018}.
\newblock


\bibitem[Redmon(2016)]%
        {redmon2016you}
\bibfield{author}{\bibinfo{person}{J Redmon}.} \bibinfo{year}{2016}\natexlab{}.
\newblock \showarticletitle{You only look once: Unified, real-time object detection}. In \bibinfo{booktitle}{\emph{Proceedings of the IEEE conference on computer vision and pattern recognition}}.
\newblock


\bibitem[Ren and O'Neill(2013)]%
        {Ren20133DSW}
\bibfield{author}{\bibinfo{person}{Gang Ren} {and} \bibinfo{person}{Eamonn O'Neill}.} \bibinfo{year}{2013}\natexlab{}.
\newblock \showarticletitle{3D selection with freehand gesture}.
\newblock \bibinfo{journal}{\emph{Comput. Graph.}}  \bibinfo{volume}{37} (\bibinfo{year}{2013}), \bibinfo{pages}{101--120}.
\newblock
\urldef\tempurl%
\url{https://api.semanticscholar.org/CorpusID:27651313}
\showURL{%
\tempurl}


\bibitem[Ren et~al\mbox{.}(2016)]%
        {ren2016faster}
\bibfield{author}{\bibinfo{person}{Shaoqing Ren}, \bibinfo{person}{Kaiming He}, \bibinfo{person}{Ross Girshick}, {and} \bibinfo{person}{Jian Sun}.} \bibinfo{year}{2016}\natexlab{}.
\newblock \showarticletitle{Faster R-CNN: Towards real-time object detection with region proposal networks}.
\newblock \bibinfo{journal}{\emph{IEEE transactions on pattern analysis and machine intelligence}} \bibinfo{volume}{39}, \bibinfo{number}{6} (\bibinfo{year}{2016}), \bibinfo{pages}{1137--1149}.
\newblock


\bibitem[Romaniak et~al\mbox{.}(2020)]%
        {Romaniak2020NimbleMI}
\bibfield{author}{\bibinfo{person}{Yevhen Romaniak}, \bibinfo{person}{Anastasiia Smielova}, \bibinfo{person}{Yevhenii Yakishyn}, \bibinfo{person}{Valerii Dziubliuk}, \bibinfo{person}{Mykhailo Zlotnyk}, {and} \bibinfo{person}{Oleksandr Viatchaninov}.} \bibinfo{year}{2020}\natexlab{}.
\newblock \showarticletitle{Nimble: Mobile Interface for a Visual Question Answering Augmented by Gestures}.
\newblock \bibinfo{journal}{\emph{Adjunct Proceedings of the 33rd Annual ACM Symposium on User Interface Software and Technology}} (\bibinfo{year}{2020}).
\newblock
\urldef\tempurl%
\url{https://api.semanticscholar.org/CorpusID:222800232}
\showURL{%
\tempurl}


\bibitem[Ruofei et~al\mbox{.}(2022)]%
        {Ruofei2022OpportunisticIF}
\bibfield{author}{\bibinfo{person}{Du Ruofei}, \bibinfo{person}{Alex Olwal}, \bibinfo{person}{Mathieu~Le Goc}, \bibinfo{person}{WU Shengzhi}, \bibinfo{person}{Danhang Tang}, \bibinfo{person}{Yinda}, \bibinfo{person}{Zhang}, \bibinfo{person}{Jun Zhang}, \bibinfo{person}{David~Joseph Tan}, {and} \bibinfo{person}{Federico Tombari}.} \bibinfo{year}{2022}\natexlab{}.
\newblock \showarticletitle{Opportunistic Interfaces for Augmented Reality: Transforming Everyday Objects into Tangible 6DoF Interfaces Using Ad hoc UI}.
\newblock \bibinfo{journal}{\emph{CHI Conference on Human Factors in Computing Systems Extended Abstracts}} (\bibinfo{year}{2022}).
\newblock
\urldef\tempurl%
\url{https://api.semanticscholar.org/CorpusID:248259431}
\showURL{%
\tempurl}


\bibitem[Schweigert et~al\mbox{.}(2019)]%
        {schweigert2019eyepointing}
\bibfield{author}{\bibinfo{person}{Robin Schweigert}, \bibinfo{person}{Valentin Schwind}, {and} \bibinfo{person}{Sven Mayer}.} \bibinfo{year}{2019}\natexlab{}.
\newblock \showarticletitle{Eyepointing: A gaze-based selection technique}.
\newblock In \bibinfo{booktitle}{\emph{Proceedings of Mensch und Computer 2019}}. \bibinfo{pages}{719--723}.
\newblock


\bibitem[Seaborn et~al\mbox{.}(2021)]%
        {seaborn2021voice}
\bibfield{author}{\bibinfo{person}{Katie Seaborn}, \bibinfo{person}{Norihisa~P Miyake}, \bibinfo{person}{Peter Pennefather}, {and} \bibinfo{person}{Mihoko Otake-Matsuura}.} \bibinfo{year}{2021}\natexlab{}.
\newblock \showarticletitle{Voice in human--agent interaction: A survey}.
\newblock \bibinfo{journal}{\emph{ACM Computing Surveys (CSUR)}} \bibinfo{volume}{54}, \bibinfo{number}{4} (\bibinfo{year}{2021}), \bibinfo{pages}{1--43}.
\newblock


\bibitem[Sendhilnathan et~al\mbox{.}(2024)]%
        {sendhilnathan2024implicit}
\bibfield{author}{\bibinfo{person}{Naveen Sendhilnathan}, \bibinfo{person}{Ajoy~S Fernandes}, \bibinfo{person}{Michael~J Proulx}, {and} \bibinfo{person}{Tanya~R Jonker}.} \bibinfo{year}{2024}\natexlab{}.
\newblock \showarticletitle{Implicit gaze research for XR systems}.
\newblock \bibinfo{journal}{\emph{arXiv preprint arXiv:2405.13878}} (\bibinfo{year}{2024}).
\newblock


\bibitem[Shi et~al\mbox{.}(2024)]%
        {Shi2024CasualGazeTM}
\bibfield{author}{\bibinfo{person}{Yingtian Shi}, \bibinfo{person}{Yukang Yan}, \bibinfo{person}{Zisu Li}, \bibinfo{person}{Chen Liang}, \bibinfo{person}{Yuntao Wang}, \bibinfo{person}{Chun Yu}, {and} \bibinfo{person}{Yuanchun Shi}.} \bibinfo{year}{2024}\natexlab{}.
\newblock \showarticletitle{CasualGaze: Towards Modeling and Recognizing Casual Gaze Behavior for Efficient Gaze-based Object Selection}.
\newblock
\urldef\tempurl%
\url{https://api.semanticscholar.org/CorpusID:271947351}
\showURL{%
\tempurl}


\bibitem[Sibert and Jacob(2000)]%
        {sibert2000evaluation}
\bibfield{author}{\bibinfo{person}{Linda~E Sibert} {and} \bibinfo{person}{Robert~JK Jacob}.} \bibinfo{year}{2000}\natexlab{}.
\newblock \showarticletitle{Evaluation of eye gaze interaction}. In \bibinfo{booktitle}{\emph{Proceedings of the SIGCHI conference on Human Factors in Computing Systems}}. \bibinfo{pages}{281--288}.
\newblock


\bibitem[Sidenmark et~al\mbox{.}(2023)]%
        {sidenmark2023vergence}
\bibfield{author}{\bibinfo{person}{Ludwig Sidenmark}, \bibinfo{person}{Christopher Clarke}, \bibinfo{person}{Joshua Newn}, \bibinfo{person}{Mathias~N Lystb{\ae}k}, \bibinfo{person}{Ken Pfeuffer}, {and} \bibinfo{person}{Hans Gellersen}.} \bibinfo{year}{2023}\natexlab{}.
\newblock \showarticletitle{Vergence matching: Inferring attention to objects in 3d environments for gaze-assisted selection}. In \bibinfo{booktitle}{\emph{Proceedings of the 2023 CHI Conference on Human Factors in Computing Systems}}. \bibinfo{pages}{1--15}.
\newblock


\bibitem[Sidenmark et~al\mbox{.}(2020a)]%
        {sidenmark2020outline}
\bibfield{author}{\bibinfo{person}{Ludwig Sidenmark}, \bibinfo{person}{Christopher Clarke}, \bibinfo{person}{Xuesong Zhang}, \bibinfo{person}{Jenny Phu}, {and} \bibinfo{person}{Hans Gellersen}.} \bibinfo{year}{2020}\natexlab{a}.
\newblock \showarticletitle{Outline pursuits: Gaze-assisted selection of occluded objects in virtual reality}. In \bibinfo{booktitle}{\emph{Proceedings of the 2020 chi conference on human factors in computing systems}}. \bibinfo{pages}{1--13}.
\newblock


\bibitem[Sidenmark and Gellersen(2019)]%
        {sidenmark2019eye}
\bibfield{author}{\bibinfo{person}{Ludwig Sidenmark} {and} \bibinfo{person}{Hans Gellersen}.} \bibinfo{year}{2019}\natexlab{}.
\newblock \showarticletitle{Eye\&head: Synergetic eye and head movement for gaze pointing and selection}. In \bibinfo{booktitle}{\emph{Proceedings of the 32nd annual ACM symposium on user interface software and technology}}. \bibinfo{pages}{1161--1174}.
\newblock


\bibitem[Sidenmark et~al\mbox{.}(2020b)]%
        {sidenmark2020bimodalgaze}
\bibfield{author}{\bibinfo{person}{Ludwig Sidenmark}, \bibinfo{person}{Diako Mardanbegi}, \bibinfo{person}{Argenis~Ramirez Gomez}, \bibinfo{person}{Christopher Clarke}, {and} \bibinfo{person}{Hans Gellersen}.} \bibinfo{year}{2020}\natexlab{b}.
\newblock \showarticletitle{Bimodalgaze: Seamlessly refined pointing with gaze and filtered gestural head movement}. In \bibinfo{booktitle}{\emph{ACM Symposium on Eye Tracking Research and Applications}}. \bibinfo{pages}{1--9}.
\newblock


\bibitem[Sidenmark et~al\mbox{.}(2024)]%
        {Sidenmark2024ConeBubbleEC}
\bibfield{author}{\bibinfo{person}{Ludwig Sidenmark}, \bibinfo{person}{Zibo Sun}, {and} \bibinfo{person}{Hans Gellersen}.} \bibinfo{year}{2024}\natexlab{}.
\newblock \showarticletitle{Cone\&Bubble: Evaluating Combinations of Gaze, Head and Hand Pointing for Target Selection in Dense 3D Environments}.
\newblock \bibinfo{journal}{\emph{2024 IEEE Conference on Virtual Reality and 3D User Interfaces Abstracts and Workshops (VRW)}} (\bibinfo{year}{2024}), \bibinfo{pages}{642--649}.
\newblock
\urldef\tempurl%
\url{https://api.semanticscholar.org/CorpusID:270097765}
\showURL{%
\tempurl}


\bibitem[Sonoda et~al\mbox{.}(2024)]%
        {Sonoda2024DiagnosticPO}
\bibfield{author}{\bibinfo{person}{Yuki Sonoda}, \bibinfo{person}{Ryo Kurokawa}, \bibinfo{person}{Yuta Nakamura}, \bibinfo{person}{Jun Kanzawa}, \bibinfo{person}{Mariko Kurokawa}, \bibinfo{person}{Yuji Ohizumi}, \bibinfo{person}{Wataru Gonoi}, {and} \bibinfo{person}{Osamu Abe}.} \bibinfo{year}{2024}\natexlab{}.
\newblock \showarticletitle{Diagnostic performances of GPT-4o, Claude 3 Opus, and Gemini 1.5 Pro in "Diagnosis Please" cases.}
\newblock \bibinfo{journal}{\emph{Japanese journal of radiology}} (\bibinfo{year}{2024}).
\newblock
\urldef\tempurl%
\url{https://api.semanticscholar.org/CorpusID:270922496}
\showURL{%
\tempurl}


\bibitem[Suzuki et~al\mbox{.}(2020)]%
        {Suzuki2020RealitySketchER}
\bibfield{author}{\bibinfo{person}{Ryo Suzuki}, \bibinfo{person}{Rubaiat~Habib Kazi}, \bibinfo{person}{Li-Yi Wei}, \bibinfo{person}{Stephen DiVerdi}, \bibinfo{person}{Wilmot Li}, {and} \bibinfo{person}{Daniel Leithinger}.} \bibinfo{year}{2020}\natexlab{}.
\newblock \showarticletitle{RealitySketch: Embedding Responsive Graphics and Visualizations in AR through Dynamic Sketching}.
\newblock \bibinfo{journal}{\emph{Proceedings of the 33rd Annual ACM Symposium on User Interface Software and Technology}} (\bibinfo{year}{2020}).
\newblock
\urldef\tempurl%
\url{https://api.semanticscholar.org/CorpusID:221186940}
\showURL{%
\tempurl}


\bibitem[Tanriverdi and Jacob(2000)]%
        {tanriverdi2000interacting}
\bibfield{author}{\bibinfo{person}{Vildan Tanriverdi} {and} \bibinfo{person}{Robert~JK Jacob}.} \bibinfo{year}{2000}\natexlab{}.
\newblock \showarticletitle{Interacting with eye movements in virtual environments}. In \bibinfo{booktitle}{\emph{Proceedings of the SIGCHI conference on Human Factors in Computing Systems}}. \bibinfo{pages}{265--272}.
\newblock


\bibitem[Toyama et~al\mbox{.}(2014)]%
        {toyama2014natural}
\bibfield{author}{\bibinfo{person}{Takumi Toyama}, \bibinfo{person}{Jason Orlosky}, \bibinfo{person}{Daniel Sonntag}, {and} \bibinfo{person}{Kiyoshi Kiyokawa}.} \bibinfo{year}{2014}\natexlab{}.
\newblock \showarticletitle{A natural interface for multi-focal plane head mounted displays using 3D gaze}. In \bibinfo{booktitle}{\emph{Proceedings of the 2014 International Working Conference on Advanced Visual Interfaces}}. \bibinfo{pages}{25--32}.
\newblock


\bibitem[T{\"u}t{\"u}nc{\"u} et~al\mbox{.}(2025)]%
        {tutuncu2025handover}
\bibfield{author}{\bibinfo{person}{Esen~K T{\"u}t{\"u}nc{\"u}}, \bibinfo{person}{Mar Gonzalez-Franco}, {and} \bibinfo{person}{Eric~J Gonzalez}.} \bibinfo{year}{2025}\natexlab{}.
\newblock \showarticletitle{HandOver: Enabling Precise Selection \& Manipulation of 3D Objects with Mouse and Hand Tracking}. In \bibinfo{booktitle}{\emph{Proceedings of the 38th Annual ACM Symposium on User Interface Software and Technology}}. \bibinfo{pages}{1--11}.
\newblock


\bibitem[Vidal et~al\mbox{.}(2013)]%
        {vidal2013pursuits}
\bibfield{author}{\bibinfo{person}{M{\'e}lodie Vidal}, \bibinfo{person}{Andreas Bulling}, {and} \bibinfo{person}{Hans Gellersen}.} \bibinfo{year}{2013}\natexlab{}.
\newblock \showarticletitle{Pursuits: spontaneous interaction with displays based on smooth pursuit eye movement and moving targets}. In \bibinfo{booktitle}{\emph{Proceedings of the 2013 ACM international joint conference on Pervasive and ubiquitous computing}}. \bibinfo{pages}{439--448}.
\newblock


\bibitem[Wang et~al\mbox{.}(2023a)]%
        {Wang2023GazeSAMII}
\bibfield{author}{\bibinfo{person}{Bin Wang}, \bibinfo{person}{Armstrong Aboah}, \bibinfo{person}{Zheyuan Zhang}, \bibinfo{person}{Hongyi Pan}, {and} \bibinfo{person}{Ulas Bagci}.} \bibinfo{year}{2023}\natexlab{a}.
\newblock \showarticletitle{GazeSAM: Interactive Image Segmentation with Eye Gaze and Segment Anything Model}. In \bibinfo{booktitle}{\emph{Gaze Meets ML}}.
\newblock
\urldef\tempurl%
\url{https://api.semanticscholar.org/CorpusID:269648415}
\showURL{%
\tempurl}


\bibitem[Wang et~al\mbox{.}(2023b)]%
        {wang2023detecting}
\bibfield{author}{\bibinfo{person}{Zhenyu Wang}, \bibinfo{person}{Yali Li}, \bibinfo{person}{Xi Chen}, \bibinfo{person}{Ser-Nam Lim}, \bibinfo{person}{Antonio Torralba}, \bibinfo{person}{Hengshuang Zhao}, {and} \bibinfo{person}{Shengjin Wang}.} \bibinfo{year}{2023}\natexlab{b}.
\newblock \showarticletitle{Detecting everything in the open world: Towards universal object detection}. In \bibinfo{booktitle}{\emph{Proceedings of the IEEE/CVF Conference on Computer Vision and Pattern Recognition}}. \bibinfo{pages}{11433--11443}.
\newblock


\bibitem[Wei et~al\mbox{.}(2023)]%
        {wei2023predicting}
\bibfield{author}{\bibinfo{person}{Yushi Wei}, \bibinfo{person}{Rongkai Shi}, \bibinfo{person}{Difeng Yu}, \bibinfo{person}{Yihong Wang}, \bibinfo{person}{Yue Li}, \bibinfo{person}{Lingyun Yu}, {and} \bibinfo{person}{Hai-Ning Liang}.} \bibinfo{year}{2023}\natexlab{}.
\newblock \showarticletitle{Predicting gaze-based target selection in augmented reality headsets based on eye and head endpoint distributions}. In \bibinfo{booktitle}{\emph{Proceedings of the 2023 CHI Conference on Human Factors in Computing Systems}}. \bibinfo{pages}{1--14}.
\newblock


\bibitem[Wu et~al\mbox{.}(2020)]%
        {Wu2020PhraseCutLI}
\bibfield{author}{\bibinfo{person}{Chenyun Wu}, \bibinfo{person}{Zhe Lin}, \bibinfo{person}{Scott~D. Cohen}, \bibinfo{person}{Trung Bui}, {and} \bibinfo{person}{Subhransu Maji}.} \bibinfo{year}{2020}\natexlab{}.
\newblock \showarticletitle{PhraseCut: Language-Based Image Segmentation in the Wild}.
\newblock \bibinfo{journal}{\emph{2020 IEEE/CVF Conference on Computer Vision and Pattern Recognition (CVPR)}} (\bibinfo{year}{2020}), \bibinfo{pages}{10213--10222}.
\newblock
\urldef\tempurl%
\url{https://api.semanticscholar.org/CorpusID:218551213}
\showURL{%
\tempurl}


\bibitem[Xiong et~al\mbox{.}(2023)]%
        {Xiong2023EfficientSAMLM}
\bibfield{author}{\bibinfo{person}{Yunyang Xiong}, \bibinfo{person}{Bala Varadarajan}, \bibinfo{person}{Lemeng Wu}, \bibinfo{person}{Xiaoyu Xiang}, \bibinfo{person}{Fanyi Xiao}, \bibinfo{person}{Chenchen Zhu}, \bibinfo{person}{Xiaoliang Dai}, \bibinfo{person}{Dilin Wang}, \bibinfo{person}{Fei Sun}, \bibinfo{person}{Forrest~N. Iandola}, \bibinfo{person}{Raghuraman Krishnamoorthi}, {and} \bibinfo{person}{Vikas Chandra}.} \bibinfo{year}{2023}\natexlab{}.
\newblock \showarticletitle{EfficientSAM: Leveraged Masked Image Pretraining for Efficient Segment Anything}.
\newblock \bibinfo{journal}{\emph{ArXiv}}  \bibinfo{volume}{abs/2312.00863} (\bibinfo{year}{2023}).
\newblock
\urldef\tempurl%
\url{https://api.semanticscholar.org/CorpusID:265608780}
\showURL{%
\tempurl}


\bibitem[Xu et~al\mbox{.}(2016)]%
        {xu2016deep}
\bibfield{author}{\bibinfo{person}{Ning Xu}, \bibinfo{person}{Brian Price}, \bibinfo{person}{Scott Cohen}, \bibinfo{person}{Jimei Yang}, {and} \bibinfo{person}{Thomas~S Huang}.} \bibinfo{year}{2016}\natexlab{}.
\newblock \showarticletitle{Deep interactive object selection}. In \bibinfo{booktitle}{\emph{Proceedings of the IEEE conference on computer vision and pattern recognition}}. \bibinfo{pages}{373--381}.
\newblock


\bibitem[Xu et~al\mbox{.}(2023)]%
        {Xu2023BridgingVA}
\bibfield{author}{\bibinfo{person}{Zunnan Xu}, \bibinfo{person}{Zhihong Chen}, \bibinfo{person}{Yong Zhang}, \bibinfo{person}{Yibing Song}, \bibinfo{person}{Xiang Wan}, {and} \bibinfo{person}{Guanbin Li}.} \bibinfo{year}{2023}\natexlab{}.
\newblock \showarticletitle{Bridging Vision and Language Encoders: Parameter-Efficient Tuning for Referring Image Segmentation}.
\newblock \bibinfo{journal}{\emph{2023 IEEE/CVF International Conference on Computer Vision (ICCV)}} (\bibinfo{year}{2023}), \bibinfo{pages}{17457--17466}.
\newblock
\urldef\tempurl%
\url{https://api.semanticscholar.org/CorpusID:260091742}
\showURL{%
\tempurl}


\bibitem[Yang and Landay(2019)]%
        {Yang2019InfoLEDAL}
\bibfield{author}{\bibinfo{person}{Jackie Yang} {and} \bibinfo{person}{James~A. Landay}.} \bibinfo{year}{2019}\natexlab{}.
\newblock \showarticletitle{InfoLED: Augmenting LED Indicator Lights for Device Positioning and Communication}.
\newblock \bibinfo{journal}{\emph{Proceedings of the 32nd Annual ACM Symposium on User Interface Software and Technology}} (\bibinfo{year}{2019}).
\newblock
\urldef\tempurl%
\url{https://api.semanticscholar.org/CorpusID:204812192}
\showURL{%
\tempurl}


\bibitem[Zai{\c{t}}i et~al\mbox{.}(2015)]%
        {zaicti2015free}
\bibfield{author}{\bibinfo{person}{Ionu{\c{t}}-Alexandru Zai{\c{t}}i}, \bibinfo{person}{{\c{S}}tefan-Gheorghe Pentiuc}, {and} \bibinfo{person}{Radu-Daniel Vatavu}.} \bibinfo{year}{2015}\natexlab{}.
\newblock \showarticletitle{On free-hand TV control: experimental results on user-elicited gestures with Leap Motion}.
\newblock \bibinfo{journal}{\emph{Personal and Ubiquitous Computing}}  \bibinfo{volume}{19} (\bibinfo{year}{2015}), \bibinfo{pages}{821--838}.
\newblock


\bibitem[Zamora et~al\mbox{.}(2024)]%
        {CamposZamora2024MoirWidgetsHP}
\bibfield{author}{\bibinfo{person}{Daniel~Campos Zamora}, \bibinfo{person}{Mustafa~Doga Dogan}, \bibinfo{person}{Alexa~F. Siu}, \bibinfo{person}{Eunyee Koh}, {and} \bibinfo{person}{Chang Xiao}.} \bibinfo{year}{2024}\natexlab{}.
\newblock \showarticletitle{Moir{\'e}Widgets: High-Precision, Passive Tangible Interfaces via Moir{\'e} Effect}.
\newblock \bibinfo{journal}{\emph{Proceedings of the CHI Conference on Human Factors in Computing Systems}} (\bibinfo{year}{2024}).
\newblock
\urldef\tempurl%
\url{https://api.semanticscholar.org/CorpusID:269743279}
\showURL{%
\tempurl}


\bibitem[Zhang et~al\mbox{.}(2025)]%
        {zhang2025forcepinch}
\bibfield{author}{\bibinfo{person}{Chenyang Zhang}, \bibinfo{person}{Tiffany~S Ma}, \bibinfo{person}{John Andrews}, \bibinfo{person}{Eric~J Gonzalez}, \bibinfo{person}{Mar Gonzalez-Franco}, {and} \bibinfo{person}{Yalong Yang}.} \bibinfo{year}{2025}\natexlab{}.
\newblock \showarticletitle{ForcePinch: Force-Responsive Spatial Interaction for Tracking Speed Control in XR}. In \bibinfo{booktitle}{\emph{Proceedings of the 38th Annual ACM Symposium on User Interface Software and Technology}}. \bibinfo{pages}{1--16}.
\newblock


\bibitem[Zhang et~al\mbox{.}(2022)]%
        {zhang2022dino}
\bibfield{author}{\bibinfo{person}{Hao Zhang}, \bibinfo{person}{Feng Li}, \bibinfo{person}{Shilong Liu}, \bibinfo{person}{Lei Zhang}, \bibinfo{person}{Hang Su}, \bibinfo{person}{Jun Zhu}, \bibinfo{person}{Lionel~M Ni}, {and} \bibinfo{person}{Heung-Yeung Shum}.} \bibinfo{year}{2022}\natexlab{}.
\newblock \showarticletitle{Dino: Detr with improved denoising anchor boxes for end-to-end object detection}.
\newblock \bibinfo{journal}{\emph{arXiv preprint arXiv:2203.03605}} (\bibinfo{year}{2022}).
\newblock


\bibitem[Zhang et~al\mbox{.}(2023)]%
        {Zhang2023VisionLanguageMF}
\bibfield{author}{\bibinfo{person}{Jingyi Zhang}, \bibinfo{person}{Jiaxing Huang}, \bibinfo{person}{Sheng Jin}, {and} \bibinfo{person}{Shijian Lu}.} \bibinfo{year}{2023}\natexlab{}.
\newblock \showarticletitle{Vision-Language Models for Vision Tasks: A Survey}.
\newblock \bibinfo{journal}{\emph{IEEE Transactions on Pattern Analysis and Machine Intelligence}}  \bibinfo{volume}{46} (\bibinfo{year}{2023}), \bibinfo{pages}{5625--5644}.
\newblock
\urldef\tempurl%
\url{https://api.semanticscholar.org/CorpusID:257913547}
\showURL{%
\tempurl}


\bibitem[Zhao et~al\mbox{.}(2023)]%
        {zhao2023revisiting}
\bibfield{author}{\bibinfo{person}{Xiaowei Zhao}, \bibinfo{person}{Yuqing Ma}, \bibinfo{person}{Duorui Wang}, \bibinfo{person}{Yifan Shen}, \bibinfo{person}{Yixuan Qiao}, {and} \bibinfo{person}{Xianglong Liu}.} \bibinfo{year}{2023}\natexlab{}.
\newblock \showarticletitle{Revisiting open world object detection}.
\newblock \bibinfo{journal}{\emph{IEEE Transactions on Circuits and Systems for Video Technology}} (\bibinfo{year}{2023}).
\newblock


\bibitem[Zhu and Chen(2024)]%
        {zhu2024survey}
\bibfield{author}{\bibinfo{person}{Chaoyang Zhu} {and} \bibinfo{person}{Long Chen}.} \bibinfo{year}{2024}\natexlab{}.
\newblock \showarticletitle{A survey on open-vocabulary detection and segmentation: Past, present, and future}.
\newblock \bibinfo{journal}{\emph{IEEE Transactions on Pattern Analysis and Machine Intelligence}} (\bibinfo{year}{2024}).
\newblock


\bibitem[Zhu et~al\mbox{.}(2019)]%
        {zhu2019zero}
\bibfield{author}{\bibinfo{person}{Pengkai Zhu}, \bibinfo{person}{Hanxiao Wang}, {and} \bibinfo{person}{Venkatesh Saligrama}.} \bibinfo{year}{2019}\natexlab{}.
\newblock \showarticletitle{Zero shot detection}.
\newblock \bibinfo{journal}{\emph{IEEE Transactions on Circuits and Systems for Video Technology}} \bibinfo{volume}{30}, \bibinfo{number}{4} (\bibinfo{year}{2019}), \bibinfo{pages}{998--1010}.
\newblock


\bibitem[Zou et~al\mbox{.}(2023)]%
        {Zou2023SegmentEE}
\bibfield{author}{\bibinfo{person}{Xueyan Zou}, \bibinfo{person}{Jianwei Yang}, \bibinfo{person}{Hao Zhang}, \bibinfo{person}{Feng Li}, \bibinfo{person}{Linjie Li}, \bibinfo{person}{Jianfeng Gao}, {and} \bibinfo{person}{Yong~Jae Lee}.} \bibinfo{year}{2023}\natexlab{}.
\newblock \showarticletitle{Segment Everything Everywhere All at Once}.
\newblock \bibinfo{journal}{\emph{ArXiv}}  \bibinfo{volume}{abs/2304.06718} (\bibinfo{year}{2023}).
\newblock
\urldef\tempurl%
\url{https://api.semanticscholar.org/CorpusID:258108410}
\showURL{%
\tempurl}


\end{thebibliography}

\clearpage
\onecolumn\appendix\enlargethispage{15pt}
\section{Appendix}
\begin{figure*}[h]  
\vspace{-0.5cm}
    \centering
    \includegraphics[height=\textheight]{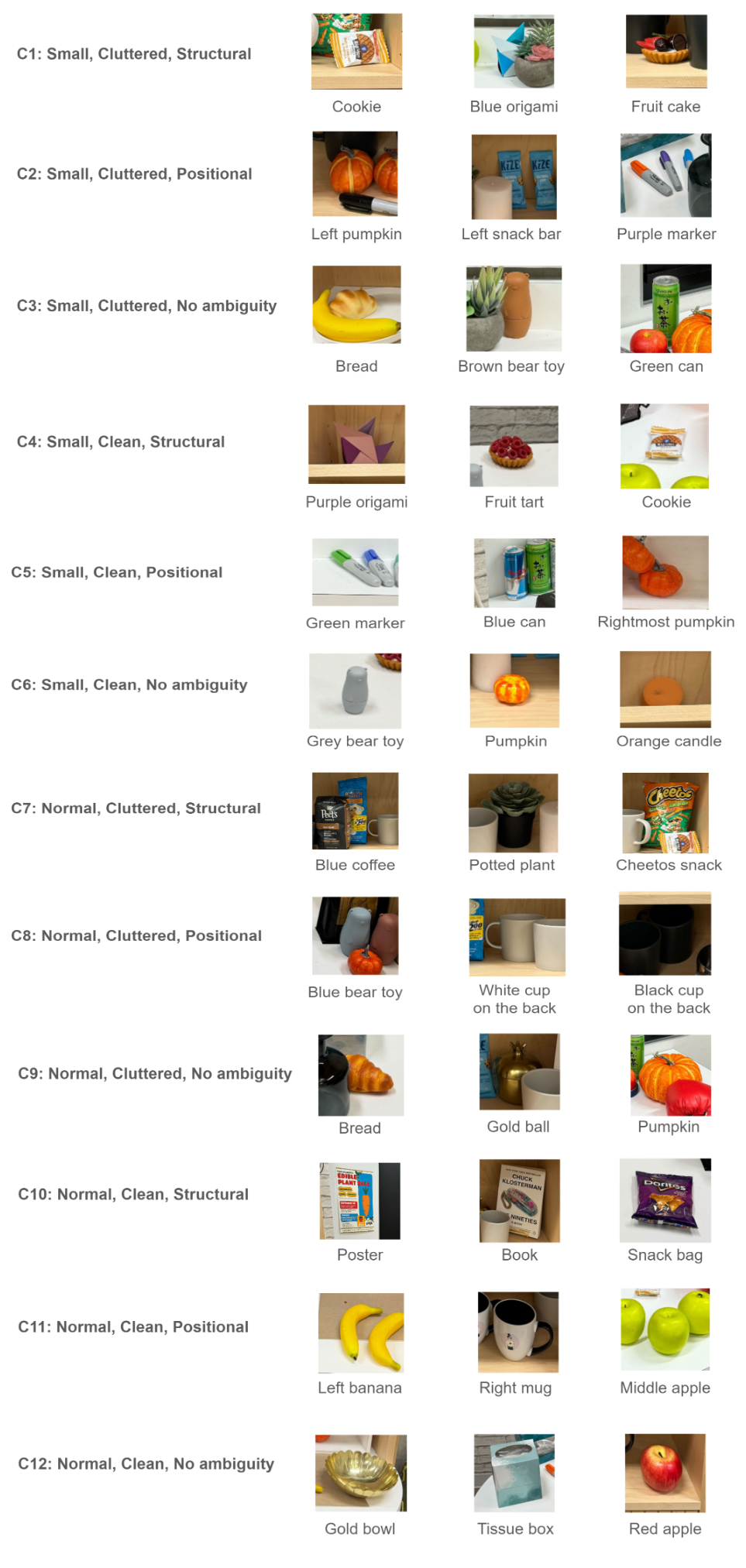}
    \caption{List of objects used in user study.}
    \label{fig:object_list}
\vspace{-4cm}
\end{figure*}

\end{document}